\documentclass[10pt]{article}
\usepackage{amsmath,amsthm,latexsym,amssymb,amsfonts,epsfig, psfrag,color,dsfont}
\usepackage[utf8]{inputenc}
\usepackage{graphicx}
\usepackage{caption}
\usepackage{subcaption}
\usepackage{sidecap}
\usepackage{wrapfig}
\usepackage{url}
\usepackage{cite}
\usepackage{tikz-cd}
\usepackage{hyperref}
\usepackage{authblk}

\makeatletter \@addtoreset{equation}{section} \makeatother

\def\be{\begin{equation}}
\def\ee{\end{equation}}
\def\ba{\begin{eqnarray}}
\def\ea{\end{eqnarray}}

\newcommand\nn{\nonumber}
\newcommand\q{\quad}
\def\Nl{{\mathchoice
{\setbox0=\hbox{$\displaystyle\rm N$}\hbox{\hbox to0pt
{\kern0.4\wd0\vrule height0.9\ht0\hss}\box0}}
{\setbox0=\hbox{$\textstyle\rm N$}\hbox{\hbox to0pt
{\kern0.4\wd0\vrule height0.9\ht0\hss}\box0}}
{\setbox0=\hbox{$\scriptstyle\rm N$}\hbox{\hbox to0pt
{\kern0.4\wd0\vrule height0.9\ht0\hss}\box0}}
{\setbox0=\hbox{$\scriptscriptstyle\rm N$}\hbox{\hbox to0pt
{\kern0.4\wd0\vrule height0.9\ht0\hss}\box0}}}}
\def\Zl{{\mathchoice
{\setbox0=\hbox{$\displaystyle\rm Z$}\hbox{\hbox to0pt
{\kern0.4\wd0\vrule height0.9\ht0\hss}\box0}}
{\setbox0=\hbox{$\textstyle\rm Z$}\hbox{\hbox to0pt
{\kern0.4\wd0\vrule height0.9\ht0\hss}\box0}}
{\setbox0=\hbox{$\scriptstyle\rm Z$}\hbox{\hbox to0pt
{\kern0.4\wd0\vrule height0.9\ht0\hss}\box0}}
{\setbox0=\hbox{$\scriptscriptstyle\rm Z$}\hbox{\hbox to0pt
{\kern0.4\wd0\vrule height0.9\ht0\hss}\box0}}}}
\def\Ql{{\mathchoice
{\setbox0=\hbox{$\displaystyle\rm Q$}\hbox{\hbox to0pt
{\kern0.4\wd0\vrule height0.9\ht0\hss}\box0}}
{\setbox0=\hbox{$\textstyle\rm Q$}\hbox{\hbox to0pt
{\kern0.4\wd0\vrule height0.9\ht0\hss}\box0}}
{\setbox0=\hbox{$\scriptstyle\rm Q$}\hbox{\hbox to0pt
{\kern0.4\wd0\vrule height0.9\ht0\hss}\box0}}
{\setbox0=\hbox{$\scriptscriptstyle\rm Q$}\hbox{\hbox to0pt
{\kern0.4\wd0\vrule height0.9\ht0\hss}\box0}}}}
\def\Rl{{\mathchoice
{\setbox0=\hbox{$\displaystyle\rm R$}\hbox{\hbox to0pt
{\kern0.4\wd0\vrule height0.9\ht0\hss}\box0}}
{\setbox0=\hbox{$\textstyle\rm R$}\hbox{\hbox to0pt
{\kern0.4\wd0\vrule height0.9\ht0\hss}\box0}}
{\setbox0=\hbox{$\scriptstyle\rm R$}\hbox{\hbox to0pt
{\kern0.4\wd0\vrule height0.9\ht0\hss}\box0}}
{\setbox0=\hbox{$\scriptscriptstyle\rm R$}\hbox{\hbox to0pt
{\kern0.4\wd0\vrule height0.9\ht0\hss}\box0}}}}
\def\Cl{{\mathchoice
{\setbox0=\hbox{$\displaystyle\rm C$}\hbox{\hbox to0pt
{\kern0.4\wd0\vrule height0.9\ht0\hss}\box0}}
{\setbox0=\hbox{$\textstyle\rm C$}\hbox{\hbox to0pt
{\kern0.4\wd0\vrule height0.9\ht0\hss}\box0}}
{\setbox0=\hbox{$\scriptstyle\rm C$}\hbox{\hbox to0pt
{\kern0.4\wd0\vrule height0.9\ht0\hss}\box0}}
{\setbox0=\hbox{$\scriptscriptstyle\rm C$}\hbox{\hbox to0pt
{\kern0.4\wd0\vrule height0.9\ht0\hss}\box0}}}}
\def\Hl{{\mathchoice
{\setbox0=\hbox{$\displaystyle\rm H$}\hbox{\hbox to0pt
{\kern0.4\wd0\vrule height0.9\ht0\hss}\box0}}
{\setbox0=\hbox{$\textstyle\rm H$}\hbox{\hbox to0pt
{\kern0.4\wd0\vrule height0.9\ht0\hss}\box0}}
{\setbox0=\hbox{$\scriptstyle\rm H$}\hbox{\hbox to0pt
{\kern0.4\wd0\vrule height0.9\ht0\hss}\box0}}
{\setbox0=\hbox{$\scriptscriptstyle\rm H$}\hbox{\hbox to0pt
{\kern0.4\wd0\vrule height0.9\ht0\hss}\box0}}}}
\def\Ol{{\mathchoice
{\setbox0=\hbox{$\displaystyle\rm O$}\hbox{\hbox to0pt
{\kern0.4\wd0\vrule height0.9\ht0\hss}\box0}}
{\setbox0=\hbox{$\textstyle\rm O$}\hbox{\hbox to0pt
{\kern0.4\wd0\vrule height0.9\ht0\hss}\box0}}
{\setbox0=\hbox{$\scriptstyle\rm O$}\hbox{\hbox to0pt
{\kern0.4\wd0\vrule height0.9\ht0\hss}\box0}}
{\setbox0=\hbox{$\scriptscriptstyle\rm O$}\hbox{\hbox to0pt
{\kern0.4\wd0\vrule height0.9\ht0\hss}\box0}}}}
\newcommand{\cc}{\mathcal C}

\newcommand{\cg}{\mathcal G}
\newcommand{\ch}{\mathcal H}

\newcommand{\cp}{\mathcal P}
\newcommand{\cq}{\mathcal Q}

\newcommand{\cs}{\mathcal S}
\newcommand{\ct}{\mathcal T}

\newcommand{\ft}{\mathfrak{t}}

\def\nn{\nonumber}

\newcommand{\eqa}{\begin{eqnarray}}
\newcommand{\neqa}{\end{eqnarray}}

\def\la{\langle}
\def\ra{\rangle}
\newcommand{\bra}[1]{\la {#1}|}
\newcommand{\ket}[1]{|{#1}\ra}

\newcommand{\p}{\partial}

\def\f{\frac}

\usepackage{bbm}

\def\q{{\quad}}

\title{How to switch between relational quantum clocks}

\author[1,2,3]{Philipp A.\ H\"ohn\thanks{\texttt{philipp.hoehn@oist.jp}} }
\author[2]{Augustin Vanrietvelde}

\affil[1]{\small Okinawa Institute of Science and Technology Graduate University, Onna, Okinawa 904 0495, Japan}
\affil[2]{\small Institute for Quantum Optics and Quantum Information, Austrian Academy of Sciences,\newline Boltzmanngasse 3, 1090 Vienna, Austria}
\affil[3]{\small Vienna Center for Quantum Science and Technology (VCQ), Faculty of Physics, University of Vienna, Boltzmanngasse 5, 1090 Vienna, Austria}

\date{}

\begin{document}

\maketitle


\begin{abstract}
Every clock is a physical system and thereby ultimately quantum. A naturally arising question is thus how to describe time evolution relative to quantum clocks and, specifically, how the dynamics relative to different quantum clocks are related. This is a particularly pressing issue in view of the multiple choice facet of the problem of time in quantum gravity, which posits that there is no distinguished choice of internal clock in generic general relativistic systems and that different choices lead to inequivalent quantum theories.
Exploiting a recent unifying approach to switching quantum reference systems \cite{Vanrietvelde:2018pgb,Vanrietvelde:2018dit}, we exhibit a systematic method for switching between different clock choices in the quantum theory. We illustrate it by means of the parametrized particle, which, like gravity, features a Hamiltonian constraint. We explicitly switch between the quantum evolution relative to the non-relativistic time variable and that relative to the particle's position, which requires carefully regularizing the zero-modes in the so-called time-of-arrival observable. While this toy model is simple, our approach is general and, in particular, directly amenable to quantum cosmology. It proceeds by systematically linking the reduced quantum theories relative to different clock choices {\it via} the clock-choice-neutral Dirac quantized theory, in analogy to coordinate changes on a manifold. This method suggests a new perspective on the multiple choice problem, indicating that it is rather a multiple choice {\it feature} of the complete relational quantum theory, taken as the {\it conjunction} of Dirac quantized and quantum deparametrized theories. Precisely this conjunction permits one to consistently switch between different temporal reference systems, which is a prerequisite for a quantum notion of general covariance. Finally, we show that quantum uncertainties generically lead to a discontinuity in the relational dynamics when switching clocks, in contrast to the classical case.
\end{abstract}

\section{Introduction}

In non-relativistic and special relativistic physics, we are used to time $t$ being described as a non-dynamical parameter or family of parameters with respect to which dynamical degrees of freedom evolve. While such a conception of time has been extremely successful in describing experiments, it is important to remind oneself that this is an idealization and that such an `externally given' time can never actually be measured in a physical experiment. Any real experiment determining the dynamics of some quantity $Q$ is based on a clock $T$ and any clock is a physical system itself. At best, the experimenter can hope that $T$ features a simple and monotonic behavior in the external $t$. However, as shown in \cite{Unruh:1989db}, no physical clock, described by quantum theory itself, can provide a perfect measure of the abstract $t$ as it would feature either a non-vanishing probability for occasionally running backwards in $t$ or states corresponding to different clock readings that are not perfectly distinguishable. This is related to Pauli's observation \cite{pauli} that there is no observable for time in standard quantum mechanics.
 Hence, even if one wanted to determine $Q(t)$, one can only determine $Q(T)$. There is no possibility to verify that $T\propto t$; at best one could try to synchronize clocks and employ some other clock $T'$ and check $T(T')$, etc. 
 
Any physical notion of time is thus a {\it relational} one and so measured and defined by {\it physical} temporal reference systems, which we usually call clocks. In particular, we consider coincidences between dynamical systems, often synchronizing one with respect to another, when recording time evolution in practice. Owing to the universality of quantum theory, every temporal reference system is ultimately quantum in nature and so a fundamentally relational description of physics faces the task to consistently describe dynamics relative to quantum systems. 

This question arises naturally and necessarily in quantum gravity and leads to the infamous {\it problem of time} \cite{Kuchar:1991qf,Isham:1992ms,Isham:1993ha,Anderson:2017jij,Rovelli:2004tv}. Essentially, it is the problem of how to extract the dynamics from a non-perturbative quantum theory of gravity, which cannot be quantized with respect to some background (coordinates) because, owing to the diffeomorphism symmetry of general relativity, there is no background with respect to which the gravitational field evolves. This leads to the {\it a priori} seemingly timeless Wheeler-DeWitt equation from which one has to extract a dynamical interpretation, by using some dynamical quantum degrees of freedom as physical reference systems with respect to which to describe the remaining physics. This has led to the {\it relational paradigm} and a considerable amount of work on relational observables, which encode how dynamical degrees of freedom are described relative to others \cite{DeWitt:1967yk,Kuchar:1991qf,Isham:1992ms,Isham:1993ha,Brown:1994py,Anderson:2017jij,relrov,Rovelli:1990jm, Rovelli:1989jn,Rovelli:1990pi, Rovelli:2004tv, Gambini:2000ht,Rovelli:2013fga, Dittrich:2004cb, Dittrich:2005kc, Dittrich:2006ee,Dittrich:2007jx,Ashtekar:2006uz,Pons:2009cz,Kaminski:2008td,Kaminski:2009qb,Tambornino:2011vg,Domagala:2010bm,Husain:2011tk,Dittrich:2016hvj, Dittrich:2015vfa ,Bojowald:2010xp,Bojowald:2010qw, Hohn:2011us, Thiemann:2007zz,  Marolf:1994nz, Marolf:2009wp,pacosmo,Hoehn:2019owq,Hoehn:2020epv}.

There has also been a revived interest in clocks as quantum reference systems in quantum information and foundations, in particular, in the context of extending the Page-Wootters approach to a viable conditional probability interpretation beyond specialized scenarios \cite{Page:1983uc,Giovannetti:2015qha,nikolova2018relational,smith2017quantizing,mendes2018time,Boette:2018uix,Hoehn:2019owq,Hoehn:2020epv,castro-ruizTimeReferenceFrames2019} and in studying quantum and thermodynamic limitations on clocks \cite{Gambini:2006yj,Gambini:2006ph,Gambini:2008ke,Gambini:2010ut,erker2017autonomous,ruiz2017entanglement}. This has also been used to attempt a formulation of quantum mechanics in terms of physical clocks, rather than external time parameters  \cite{Gambini:2006yj,Gambini:2006ph,Gambini:2008ke,Gambini:2010ut}. 

Quantum clocks thus appear in  quantum gravity, as well as in quantum foundations and in this article, we shall study an elementary question that is pertinent to both fields.
Consider evolving two clocks, represented by variables $T$ and $Q$, relative to one another. For instance, $T$ could represent the clock in a laboratory and $Q$ the position of a freely moving  particle, which itself we could use as a clock. What is their relation? Classically, this question is, of course, in principle straightforward to address. It requires to solve the equations of motion (relative to some time parameter $t$) and the relative evolution $Q(T)$ is classically, of course, equivalent to $T(Q)$ since one could simply (at least locally) invert a solution. Classically, it is thus in principle straightforward to switch from the evolution relative to one clock to that relative to another. But how can one consistently relate and switch clocks in a completely relational quantum theory, when $\hat{T}$ and $\hat{Q}$ are operators and an external reference time $t$ is absent? That this will be a substantially harder task is already clear from the discussion surrounding the challenging time-of-arrival concept in quantum mechanics \cite{Grot:1996xu,aharonov1998measurement,muga2000arrival,Gambini:2000ht }, which aims at providing a quantum implementation of the time $T(Q)$ when a particle reaches a position $Q$.

In general, there will, of course, exist a lot of different physical systems that could serve as a clock. Which one should one pick and if one considers evolution with respect to one, how will the evolution look like relative to a different choice of clock if they are genuine quantum systems? For example, in full general relativity and quantum gravity (i.e.\ away from symmetry reductions or specialized matter content) there is no natural choice of internal clock function among the geometric or matter degrees of freedom relative to which to evolve the remaining ones. This leads to what Kucha\v{r} and Isham coined the {\it multiple choice problem} in quantum gravity and generally covariant quantum systems \cite{Kuchar:1991qf,Isham:1992ms}. The purported problem is that the various quantum theories relative to different clock choices are generally  inequivalent. Indeed, the argument is based on the observation that different clock choices are related by canonical transformations at the classical level and that, on account of the Groenewold-Van-Hove phenomenon \cite{groenewold1946principles,van1951problem} in non-linear systems, most canonical transformations cannot be represented {\it unitarily} in the quantum theory. And so Kucha\v{r} states: {\it ``The multiple
choice problem is one of an embarrassment of riches: out of many inequivalent
options, one does not know which one to select."} \cite{Kuchar:1991qf}. In a similar vein, Isham asks {\it ``[The dynamics] based on one particular
choice of internal time may give a different quantum theory from that based on
another. Which, if any, is correct?"} \cite{Isham:1992ms}.

In this article,  by means of a toy model, we shall propose the view that {\it all} these different relational dynamics based on different choices of internal clocks are correct,\footnote{Except for possibly arising pathological clock choices, e.g.\ see \cite{Hartle:1995vj}, and modulo possible factor ordering ambiguities.} and correspond to the {\it same} physics, but described relative to different temporal reference systems. The same physics can, of course, look different from different perspectives. In this light, we shall argue that the multiple choice facet is {\it not} a problem, but a {\it feature} of a completely relational quantum theory that, specifically, must admit a quantum notion of general covariance, i.e.\ consistent switches from one quantum reference system to another (of either spatial or temporal nature, see also the discussion in \cite{Hoehn:2017gst,Giacomini:2017zju,Vanrietvelde:2018pgb,Vanrietvelde:2018dit,FEC}). To back up our proposal, we thus have to provide a systematic method for switching between the quantum dynamics relative to different choices of internal clocks.

In particular, Isham asks {\it ``Can these different quantum theories be seen
to be part of an overall scheme that is covariant?"}, and states further
{\it ``It seems most unlikely that a single Hilbert space can be used for all possible choices of an internal time function."} \cite{Isham:1992ms}. As we shall illustrate by means of the parametrized particle, which in analogy to general relativity features a Hamiltonian constraint, the answer is ``yes'' and constructing the link between the different choices of internal time indeed requires a multitude of Hilbert spaces. While this toy model is very simple, we shall argue that the method is general and directly amenable to quantum cosmology and models of  quantum gravity. Indeed, the new clock change method introduced in the present article has later been developed further in \cite{pacosmo,Hoehn:2019owq,Hoehn:2020epv,castro-ruizTimeReferenceFrames2019}.

The first systematic approach to changing clocks in generally covariant quantum systems was developed in a quantum phase space language in \cite{Bojowald:2010xp,Bojowald:2010qw, Hohn:2011us}, yet restricted to the semiclassical regime. Here, we shall exploit a new unifying approach to switching quantum reference systems in quantum foundations and gravity \cite{Vanrietvelde:2018pgb,Vanrietvelde:2018dit}, which took inspiration from the semiclassical clock changes in \cite{Bojowald:2010xp,Bojowald:2010qw, Hohn:2011us} and the operational approach to quantum reference frames advocated in \cite{Giacomini:2017zju}, to put forward a systematic method for switching between the dynamics relative to different choices of quantum clocks.\footnote{In \cite{pqps} it will be shown that our full quantum method coincides with the semiclassical one in \cite{Bojowald:2010xp,Bojowald:2010qw, Hohn:2011us}, once restricted to the semiclassical regime.}
 
Given that our construction involves a quantum symmetry reduction procedure, it also touches on the relation between the two general methods for quantizing systems with constraints: (a) constrain first, then quantize (the so-called reduced method); and (b) quantize first, then constrain (the so-called Dirac method). The general conclusion in the literature is that `constraining and quantization do not commute' so that the two methods are, in general inequivalent and, specifically, produce different Hilbert spaces for the same system \cite{Ashtekar:1982wv,Ashtekar:1991hf,Schleich:1990gd,Kunstatter:1991ds,Hajicek:1990eu,Romano:1989zb,Dittrich:2016hvj,Dittrich:2015vfa,Loll:1990rx,Kaminski:2009qb,Domagala:2010bm,Giesel:2016gxq}. This has led to a debate about when one or the other would be the physically correct method to apply. In the context of the problem of time, this has led to three broad categories of approaches \cite{Kuchar:1991qf,Isham:1992ms}, which Isham characterizes as follows: 
\begin{description}
\item[Tempus ante quantum.] Choose the internal clock classically, solve the constraints, then quantize to produce something resembling a Schr\"odinger equation in internal time. This is thus the reduced method with respect to a specific choice of internal time.
\item[Tempus post quantum.] Quantize first, then impose constraints, producing a Wheeler-DeWitt type equation, and finally identify an internal time to interpret it. This is thus the Dirac method with a subsequent choice of clock.
\item[Tempus nihil est.] Construct a consistent and complete quantum theory that is fundamentally timeless (e.g.\ via the Dirac method) and attempt to recover dynamics from it in an emergent fashion.
\end{description} 

The main point of the new approach \cite{Vanrietvelde:2018pgb,Vanrietvelde:2018dit} is that it identifies the Dirac method as providing a perspective-neutral -- i.e., reference-system-neutral -- quantum theory \cite{Hoehn:2017gst}: it is a global description of all degrees of freedom prior to having chosen a (quantum) reference system relative to which the physics of the remaining degrees of freedom is described. As such, it encodes all permitted quantum reference system choices at once and this is reflected in the redundancy of the description of the physical Hilbert space induced by the gauge symmetry. The new approach then provides a novel systematic two-step quantum symmetry reduction procedure, relative to a choice of reference system, which maps the perspective-neutral Dirac quantized theory into a redundancy-free reduced quantum theory: (1) `trivialize' the constraints to the choice of reference system to single out its degrees of freedom as the redundant ones, and (2) subsequently condition on the classical gauge fixing conditions. The new approach identifies the quantum symmetry reduced theory as providing the quantum description relative to the perspectives of the associated choice of reference system. 

In simple cases, the result of applying the new quantum symmetry reduction procedure to the Dirac quantized theory (method (b)) will actually be unitarily equivalent to the quantization of a classically symmetry reduced phase space (method (a)). This is the case for the model in the present article and the ones in \cite{Vanrietvelde:2018pgb,pacosmo}. However, this will not be true in general on account of the typical inequivalence of the two quantization methods for constrained systems. We can thus rephrase this observation as `symmetry reduction and quantization do not commute in general' and this is further discussed in \cite{Vanrietvelde:2018dit,Hoehn:2019owq,Hoehn:2020epv}.

Applying this to the context of temporal reference systems in this manuscript, we characterize the gauge invariant physical Hilbert space of the Dirac method as a clock-choice-neutral, rather than timeless quantum structure. It is a global description  prior to having chosen a temporal reference system, i.e.\ clock, relative to which the quantum dynamics of the remaining degrees of freedom is described; accordingly, it encodes all (permissible) clock choices at once. Similarly, we interpret the quantum symmetry reduced theories as providing the description of the quantum dynamics described relative to a given choice of clock. We can explicitly construct the transformations that map the clock-neutral physical Hilbert space to the symmetry reduced Hilbert space relative to a given clock choice. Since this map will be invertible (not always globally), we can thereby construct the linking map from the dynamics relative to one choice of clock to that relative to another by concatenating one quantum reduction map with the inverse of the other, in analogy to a coordinate transformation on a manifold. This linking map from one clock choice to another thus proceeds via the clock-neutral physical Hilbert space just like coordinate transformations pass through the manifold. 

We conclude that {\it a complete relational quantum theory, which features a quantum notion of general covariance  requires {\emph{both}} the perspective-neutral Dirac quantized theory and the multitude of theories which are quantum symmetry reduced with respect to a choice of quantum reference system.} 

As mentioned above, for the simple models considered here and in \cite{pacosmo}, the quantum symmetry reduced theories will coincide with the quantization of classically symmetry reduced theories (method (a)). Our method thereby links tempus-ante-quantum and tempus-post-quantum strategies in simple cases. In fact, by identifying the physical Hilbert space of the Dirac method as a clock-neutral quantum structure, our method also encompasses the so-called frozen time formalism \cite{relrov,Rovelli:1990jm, Rovelli:1989jn,Rovelli:1990pi, Rovelli:2004tv}, a tempus-nihil-est strategy, according to Isham \cite{Isham:1992ms}.

Coming back to comparing clock readings, as an application of the novel quantum clock change method we show that, in contrast to the classical case, quantum uncertainties generically lead to a discontinuity in the relational dynamics when changing from one clock to another.\footnote{No such discontinuity occurs in the model later studied in \cite{pacosmo}. However, this is a special case due to a high degree of symmetry between different clock choices.} This resonates with the semiclassical observations in \cite{Bojowald:2010xp,Bojowald:2010qw, Hohn:2011us} and can also be viewed as yielding a temporal reference system dependence or relativity of quantum clock comparisons.

Our general method is not in conflict with the Groenewold-Van-Hove phenomenon: while the reduced quantum theories based on different clock choices are linked, they will not always be unitarily equivalent. In generic systems, these linking maps will be isometries but, just like coordinate changes on a manifold, not globally valid (on the physical Hilbert space), in analogy to the Gribov problem. Hence, they will not necessarily provide a global isometric equivalence between the various Hilbert spaces. Indeed, the quantum symmetry reduction will not provide a globally valid quantum theory if it projects onto classical gauge choices that are not valid globally. This is not a fundamental problem; it reflects the fact that global `perspectives' in physics are generally unavailable, but that one can still have valid non-global descriptions of the physics that, in some cases, can even be `patched up' to global ones \cite{Bojowald:2010xp,Bojowald:2010qw, Hohn:2011us,Vanrietvelde:2018dit }.  

This is especially relevant in view of the so-called {\it global problem of time}: a generic general relativistic system is devoid of internal clock functions that always run `forward' \cite{Kuchar:1991qf,Isham:1992ms,Anderson:2017jij,Bojowald:2010xp,Bojowald:2010qw, Hohn:2011us,Dittrich:2016hvj,Dittrich:2015vfa,Hajicek:1986ky, Hajicek:1994py,Hajicek:1995en,Hajicek:1988he,Schon:1989pe,Hajicek:1989ex,Rovelli:1990jm}. In such systems, any clock choice will thus necessarily produce a reduced theory that is not globally valid, as any clock will eventually encounter a turning point, which means that the relational dynamics with respect to it will eventually become non-unitary. Especially in such a context it may be essential to have a systematic method for switching temporal reference systems at hand, so one can, at least in some cases, sidestep non-unitarity by only employing a clock choice in a transient manner and switching  to another before it becomes pathological \cite{Bojowald:2010xp,Bojowald:2010qw, Hohn:2011us}. (However, see \cite{Hohn:2011us,Dittrich:2016hvj,Dittrich:2015vfa} for challenges to this strategy in the presence of chaos.)

We note that changing quantum clocks was also considered in \cite{Malkiewicz:2014fja, Malkiewicz:2015fqa,Malkiewicz:2016hjr,Malkiewicz:2017cuw} at the level of reduced quantization and for a restricted set of clock choices. Some interesting physical consequences were studied, but the relation to Dirac quantization remains unclear and so the method in \cite{Malkiewicz:2014fja, Malkiewicz:2015fqa,Malkiewicz:2016hjr,Malkiewicz:2017cuw} did not provide a comprehensive picture for general quantum covariance.
 It is an open and interesting question what the relation, if any, of that method is to the unifying one proposed here.

The remainder of this article is organized as follows. In sec.\ \ref{sec_cl}, we revisit the classical parametrized non-relativistic particle, however, from a novel angle that will prepare the new quantum method. In particular, we show how to switch from the relational dynamics in the non-relativistic time variable to the time-of-arrival dynamics relative to the particle's position, by mapping the correspondingly gauge-fixed reduced phase spaces to one another. This requires to separate left from right moving solutions. In sec.\ \ref{sec_q}, we then quantize this method, by first providing the reduced quantum theories relative to the two clock choices in secs.\ \ref{sec_qttime} and \ref{sec_qqtime}, and subsequently linking them via the clock-neutral Dirac quantized theory in secs.\ \ref{sec_Dirac}--\ref{sec_qclockswitch}. The main challenge is the relational dynamics relative to the particle's position, which requires to carefully regularize and quantize the time-of-arrival function \cite{Grot:1996xu,aharonov1998measurement,muga2000arrival,Gambini:2000ht } to a self-adjoint relational observable in both the Dirac and reduced method. Remarkably, the new quantum reduction method consistently maps the regularized observables from the Dirac quantized theory to the correctly regularized ones of the reduced theories. As an application of the new method, we exhibit in sec.~\ref{sec_qrelativity} a generic discontinuity in the relational evolution when switching clocks, which can also be interpreted as a `quantum relativity' of comparing clock readings. Finally, we conclude with an outlook in sec.\ \ref{sec_conc}. Many technical details have been moved to various appendices.

\section{Revisiting the classical parametrized non-relativistic particle}\label{sec_cl}

It is instructive to illustrate the new method of relational clock changes first in a very simple system, namely the parametrized Newtonian free particle, which has been used many times before to illustrate the basic ideas underlying the paradigm of relational dynamics (e.g., see \cite{Rovelli:2004tv, Tambornino:2011vg, bianca-lecnotes, Henneaux:1992ig}). However, as we shall see, this simple model still features a surprisingly non-trivial behavior, once employing its position as a relational clock. We begin by revisiting this toy model from a somewhat novel classical perspective to prepare for the subsequent new quantum method of clock changes.

Given that this is a standard toy model, we shall directly jump into its canonical formulation. However, for the unacquainted reader we provide its derivation from a reparametrization-invariant action principle in appendix \ref{app_nrpart} to keep this article self-contained.

The parametrized free particle is described by two canonical pairs $(q,p), (t,p_t)$ on a four-dimensional phase space $\mathbb{R}^4$. Here, $(t,p_t)$ describes the time coordinate, which has been promoted to a dynamical variable, and its conjugate momentum. Setting, for convenience, the mass to $m=1/2$ so we do not have to carry these factors around, the reparametrization symmetry of its dynamics produces a single {\it Hamiltonian constraint} (see appendix \ref{app_nrpart})
\ba
C_H=p_t+p^2\approx0\,,\label{CH}
\ea
i.e., the Hamiltonian for this system has to vanish on solutions. This defines a three-dimensional constraint surface $\cc$ in phase space. The symbol $\approx$ denotes a weak equality, i.e.\ an equality that only holds on this constraint surface \cite{Dirac,Henneaux:1992ig}.
 
The flow parameter $s$ along the orbits generated by the Hamiltonian constraint $C_H$ in $\cc$ constitutes an unphysical `time coordinate' $s$ that is itself not dynamical. 
It rather assumes the role of a gauge parameter, given that the system features a reparametrization symmetry. We can therefore not directly interpret the changes generated by $C_H$, namely the equations of motion\footnote{We have set the lapse to $N=1$, see appendix \ref{app_nrpart}.}
\ba
t'&=&\{t,C_H\}\,=1\,,\q\q\q\q \;\, p_t'=\{p_t,C_H\}=0\,,\nn\\
q'&=&\{q,C_H\}=2p\,,\q\q\q\q p'=\,\{p,C_H\}\,\,=0\,,\label{nreom}
\ea
as physical motion, where a $'$ denotes differentiation with respect to $s$. Indeed, owing to the reparametrization symmetry of the system, any physical information must be reparametrization-invariant. Since reparametrization-invariant information is encoded in functions $O$ on $\cc$ that Poisson-commute with $C_H$, $\{O,C_H\}\approx0$ -- the Dirac observables of this system -- we have to encode the gauge invariant information about the dynamics in constants of motion. Clearly, $p_t$ and $p$ are both (dependent) Dirac observables. In order to also encode invariant dynamical information about $t$ and $q$, we have to resort to {\it evolving constants of motion}, i.e.\ relational Dirac observables \cite{relrov,Rovelli:1990jm, Rovelli:1989jn,Rovelli:1990pi, Rovelli:2004tv, Gambini:2000ht,Rovelli:2013fga, Dittrich:2004cb, Dittrich:2005kc, Dittrich:2006ee,Dittrich:2007jx,Ashtekar:2006uz,Pons:2009cz,Kaminski:2008td,Kaminski:2009qb,Tambornino:2011vg,Domagala:2010bm,Husain:2011tk,Dittrich:2016hvj, Dittrich:2015vfa ,Bojowald:2010xp,Bojowald:2010qw, Hohn:2011us }, as we shall explain shortly.

The equations of motion are solved by $t(s)=s+t_0$ and $q(s)=2ps+q_0$ and so it is clear that
\ba
Q=q(s)-2p\,t(s)=q_0-2p\,t_0\label{Dirac1}
\ea
is a constant of motion too. From this Dirac observable we can construct {\it relational} Dirac observables in multiple ways. 
Indeed, at this stage, we have to make an additional choice that is not dictated by the system or formalism itself: we have to divide the dynamical degrees of freedom of the system into evolving degrees of freedom and a temporal reference system, henceforth also called an internal, or relational `clock'. This clock will constitute the `time standard' relative to which we describe the dynamics of the remaining degrees of freedom. 

Since the structure introduced thus far, and in particular the constraint surface $\cc$, encodes {\it all} these choices at once, we can safely interpret it as a clock-choice-neutral super structure (see also \cite{Bojowald:2010qw} for the semiclassical analog). As such, it does not have an immediate physical interpretation because we simply have not yet chosen a `reference frame' from which to describe the remaining physics. This is also reflected in the reparametrization-symmetry related redundancy in the description of $\cc$. For example, the two Dirac observables $p,p_t$ are dependent by (\ref{CH}), but of course the system does not tell us which of the two to regard as the redundant one. Furthermore, given that the constraint $C_H$ generates a one-dimensional gauge orbit, the reduced phase space will be two-dimensional. 
Hence, we will only have two independent gauge invariant degrees of freedom and there are many ways in choosing them and, accordingly, in fixing and removing redundant degrees of freedom. 

In particular, after we choose a specific temporal reference system with respect to which we describe the remaining dynamics, we no longer wish to consider its degrees of freedom as dynamical variables. Choosing a reference system means describing all physics relative to it and describing the reference system relative to itself does not yield dynamical information.
Instead, as we shall see, upon gauge fixing, the temporal reference will assume the role of a non-dynamical evolution {\it parameter} on a gauge-fixed reduced phase space, eliminating redundant dynamical information. That is, the gauge-fixed reduced phase spaces will be interpreted as encoding the physics described relative to a particular temporal reference system.
As such, it is these reduced phase spaces, which, in fact, admit a direct physical interpretation, in contrast to the clock-choice-neutral super structure $\cc$ (which also is not a phase space).

This extends the interpretation proposed in \cite{Vanrietvelde:2018pgb, Vanrietvelde:2018dit} for spatial reference frames to the temporal case. Indeed, in \cite{Vanrietvelde:2018pgb, Vanrietvelde:2018dit} it was put forward to interpret the constraint surface (of a first class system) as a perspective-neutral structure, which contains all reference frame choices at once. By  contrast,  it was advocated that ``jumping into the perspective of a specific (spatial) reference frame", from which to describe the remaining physics, corresponds to a gauge choice and restricting to the associated reduced phase. Therefore, for the spatial frames of \cite{Vanrietvelde:2018pgb, Vanrietvelde:2018dit} it is also precisely these gauge-fixed reduced phase spaces, which admit a direct interpretation as the physics seen from a specific perspective.

That is, in both the spatial and temporal case, leaving the reference-system-neutral grounds is tantamount to fixing and removing the redundancy. We shall detail this now for the temporal case which will amount to a deparametrization of the model.

\subsection{Relational evolution in $t$-time}\label{sec_clttime}

Let us firstly choose $t$ as the internal clock. Denote the relational Dirac observables describing evolution of $q,p$ with respect to $t$ by $Q(\tau),P(\tau)$. They are defined as the cooincidences of $q,p$ with $t$, when $t$ reads the value $\tau$, i.e.\ $Q(\tau)=q(s)\left|_{t(s)=\tau}\right.$ and likewise for $P$. That is, $Q(\tau)$ encodes the question: ``what is the position $q$ of the particle when the clock $t$ reads $\tau$?". Using (\ref{Dirac1}), we find
\ba
Q(\tau)=Q+2p\,\tau=2p(\tau-t_0)+q_0\,,\q\q\q\q\q P(\tau)=p\,.\label{QT}
\ea
Indeed, $Q(\tau)$ commutes with the constraint $\{Q(\tau),C_H\}=-2p+2p=0$ $\forall\, \tau$ and is thus gauge-invariant. The two relational observables also form a canonical pair $\{Q(\tau),P(\tau)\}=1$ and it is clear that describing the clock relative to itself yields just the evolution {\it parameter} and a redundant Dirac observable
\ba
T(\tau):=t(s)\left|_{t(s)=\tau}\right.=\tau\,,\q\q\q\q\q P_T(\tau):=p_t(s)\left|_{t(s)=\tau}\right.=p_t\,.\label{TT}
\ea
The relational observables (\ref{QT}, \ref{TT}) correspond to the reparametrization-invariant information gathered by `scanning' with $t=const$ slices through the constraint surface $\cc$.

Notice that, while setting $t(s)=\tau$ corresponds to a gauge-fixing of the flow of $C_H$, $Q(\tau)$ is a function on the entire constraint surface. Namely, given that $Q$ in (\ref{Dirac1}) is a constant of motion, it does not matter on which point on a fixed orbit in $\cc$ to evaluate it. $Q(\tau)$ only depends on the orbit, defined through the initial data $(q_0,t_0,p)$ ($p_t$ is redundant), and is thus a gauge-invariant extension of a gauge-fixed quantity \cite{Henneaux:1992ig,Dittrich:2004cb, Dittrich:2005kc}.
Now $\tau$ is the {\it parameter} which runs over all values that $t(s)$ can take on the given orbit. Hence, the two parameter families of Dirac observables $Q(\tau),P(\tau)$ describe the complete gauge invariant relational evolution of $q,p$ with respect to the dynamical $t$. In particular, since they are constants of motion, we can now evaluate the {\it entire} dynamics along a given orbit by restricting to any single point on it and just letting the parameter $\tau$ run.

At this stage, we can thus choose to fix a gauge, e.g., by setting $t=0$ (which due to (\ref{nreom}) intersects every orbit once and only once), to construct a reduced phase space and get rid of the redundancy of the description on $\cc$. This will constitute a deparametrization of the model (relative to $t$). The Dirac bracket \cite{Dirac,Henneaux:1992ig}, defining the inherited bracket structure on the reduced phase space, is particularly simple in this case
\ba
\{F,G\}_D=\{F,G\}-\{F,C_H\}\{t,G\}+\{F,t\}\{C_H,G\}\,,\nn
\ea
where $\{.,.\}$ denotes the Poisson bracket and $F,G$ are any functions on $\cc$. Specifically,
\ba
\{t,p_t\}_D=\{t,f(q,p)\}_D=0\,,\q\q\{p_t,f(q,p)\}_D=-\{p^2,f(q,p)\}_D\,,\q\q \{q,p\}_D=\{q,p\}=1\,,\nn
\ea
so that we can consistently discard the redundant reference system pair $(t,p_t)$ from among the dynamical variables and keep $(q,p)$ to coordinatize the gauge-fixed two-dimensional reduced phase space. We shall denote the latter by $\cp_{q|t}\simeq\mathbb{R}^2$ to highlight that it is $q$ that is evolving in $t$.

Although the gauge-fixed reduced phase space $\cp_{q|t}$ embeds into $\cc$ as the `$t=0$-slice' $\cg_{t=0}\cap\cc$, where $\cg_{t=0}$ is the gauge fixing surface $t=0$, it contains all the dynamical information thanks to the observation above. Namely, we can evaluate the two families of Dirac observables $Q(\tau),P(\tau)$ for all values of $\tau$\footnote{Recall from (\ref{TT}) that this parameter is really the gauge-invariant observable $T(\tau)$.} in this single reduced phase space, by setting $t=0$,
\ba
Q(\tau)=q+2p\,\tau\,,\q\q\q\q\q P(\tau)=p\,.\label{QT2}
\ea
Differentiating with respect to the evolution parameter $\tau$, we obtain their equations of motion in $\cp_{q|t}$
\ba
\f{\partial Q}{\p\tau}=2p=\{Q,H\}_D\,,\q\q\q\q \f{\p P}{\p \tau}=0=\{P,H\}_D\,, \nn
\ea 
which are thus generated by the {\it physical} Hamiltonian $H(\tau)=P^2(\tau)=p^2$ (and with respect to the Dirac bracket). That is, the gauge-fixed reduced phase space coincides with the phase space of the unparametrized Newtonian free particle and the evolution parameter $\tau$ takes the role of the Newtonian absolute time; the system has been deparametrized. The relational dynamics of the parametrized particle in this reduced form admits a direct physical interpretation and is, of course, entirely equivalent to the dynamics of the unparametrized particle (see also appendix \ref{app_nrpart}).


%
%
%
%

\subsection{Relational evolution in $q$-time as time of arrival}\label{sec_clqtime}

Next, we interchange the roles of $t$ and $q$, choosing the position $q$ as an internal `clock' and ask for the values of $t,p_t$ when $q$ takes the value $X$. Again, using (\ref{Dirac1}) yields the evolving constants of motion
\ba
T(X):=t(s)\left|_{q(s)=X}\right. = \f{X-Q}{2p} =t_0+ \f{X-q_0}{2p}\,,\q\q\q\q\q P_T(X):=p_t(s)\left|_{q(s)=X}\right. =p_t\,.\label{TX}
\ea
Clearly, $\{T(X),C_H\}=0$ and $\{T(X),P_T(X)\}=1$ for all $X$. In analogy to before, evolving the new `clock' relative to itself just yields the evolution {\it parameter}, $Q(X):=q(s)\left|_{q(s)=X}\right.=X$. Note that $T(X)$ is the {\it time-of-arrival function} (here as a Dirac observable) \cite{Grot:1996xu,aharonov1998measurement,muga2000arrival,Gambini:2000ht,Dittrich:2006ee, bianca-lecnotes}; it embodies the question ``what is the time $t$ when the particle reaches position $q=X$?"

Just like $Q(\tau),P(\tau)$ above, $T(X),P_T(X)$ are gauge-invariant extensions of gauge-fixed quantities. We would thus again like to evaluate the two parameter families entirely on a single gauge-fixing surface $q=const$ to fix and remove redundant degrees of freedom, in this case the pair $(q,p)$ corresponding to the new temporal reference. However, here we have to be more careful than in $t$-time and this has to do with the Dirac observable $p$, which we would now like to treat as redundant. 
\begin{itemize}
\item[(a)] As can be seen from (\ref{nreom}), $q=const$ fails to be a valid gauge fixing condition for $p=0$. When $p=0$, $q$ is constant along the orbit, while $t$ always grows monotonically. Hence, $q$ can then not be used to fix the flow of $C_H$ and is also the worst possible `clock' for resolving the evolution of $t$. Correspondingly, $T(X)$ becomes ill-defined for $p=0$. A stationary point particle is a bad clock.

For example, the $q=0$-slice $\cg_{q=0}$ covers the entire $(q=0,p=0)$-orbit and misses all other orbits with $p=0$. Thus, the intersection $\cg_{q=0}\cap\cc$ will {\it not} be equivalent to the (abstract) reduced phase space, which is the space of orbits $\cp_{\rm red}:=\cc/\sim$, where $\sim$ identifies points if they lie in the same orbit.

\item[(b)] On the gauge-fixed reduced phase space, $p$ will no longer be a variable and we have to replace it. Through (\ref{CH}) it admits two solutions in terms of the surviving $p_t$, corresponding, of course, to left and right moving solutions, which we will have to distinguish in the sequel, also in view of the subsequent quantum theories.
\end{itemize}

To cope with these issues, it is convenient to factorize the Hamiltonian constraint as follows
\ba
C_H=C_+C_-\,,\q\q\q\q\q C_\pm:=p\pm h\,,\q\q\q\q\q h:=\sqrt{-p_t}\,.\label{factorize}
\ea
Notice that $p_t\leq0$ on $\cc$, so $h$ defines a good Hamiltonian. The equations of motion can then be recast
\ba
\f{\mathrm{d}\,\cdot}{\mathrm{d}s}=\{\cdot,C_H\}=C_+\,\{\cdot,C_-\}+\{\cdot,C_+\}\,C_-\,,
\ea
and we can distinguish the following three situations for $C_H=0$:
\begin{itemize}
\item[(i)] When $C_+=0$ and $p\neq0$, we have
\ba
\f{\mathrm{d}\,\cdot}{\mathrm{d}s}=\{\cdot,C_H\}\approx -2h\,\{\cdot,C_+\}\,,
\ea
and so $C_+$ generates the dynamics in this region of $\cc$. Since $h>0$, the flow generated by $C_+$ is directed opposite to that of $C_H$ (i.e., the Hamiltonian vector fields of $C_+$ and $C_H$ point in opposite directions). Specifically, while the flow of $C_+$ always moves $q$ `forward' (i.e.\ to the right) because $\{q,C_+\}=1$, we have here $q'\approx -2h<0$ so that $C_H$ generates the left moving solutions in the region where $C_+=0$ and $p\neq0$.

\item[(ii)] When $C_-=0$ and $p\neq0$, we have
\ba
\f{\mathrm{d}\,\cdot}{\mathrm{d}s}=\{\cdot,C_H\}\approx 2h\,\{\cdot,C_-\}\,,
\ea
and so $C_-$ generates the dynamics in this region of $\cc$. Since $h>0$, the flows (and Hamiltonian vector fields) of $C_+$ and $C_H$ are aligned. Here, $q'\approx 2h>0$ so that the region where $C_-=0$ and $p\neq0$ corresponds to the solutions where the particle moves to the right.

\item[(iii)] When $p=p_t=0$, we have $C_+=C_-=0$. This is the shared boundary between the regions of $\cc$ where $C_+$ and $C_-$ vanish. Their gradients diverge here as $\mathrm{d}C_\pm=\mathrm{d}p\mp \f{1}{\sqrt{-p_t}}\,\mathrm{d}p_t$ so that $C_\pm$ fail to satisfy the standard regularity conditions \cite{Henneaux:1992ig} and we can no longer employ them as evolution generators. Notice that the original constraint $C_H$ always has a well-defined gradient $\mathrm{d}C_H=\mathrm{d}p_t+2p\,\mathrm{d}p$; indeed, the equations of motion (\ref{nreom}) are always well-defined.

\end{itemize}

Since $p=0$ causes trouble in any case ((a) and (iii)) for the construction of a gauge-fixed reduced phase space and $C_\pm$ are well-behaved for $p\neq0$, it will be more convenient to use the latter to define two sub-constraint surfaces $\cc_\pm$ via $C_\pm=0$ within $\cc$. The constraints $C_\pm$ encode the same gauge-invariant information as $C_H$ on $\cc_\pm$ because solving the dynamics generated by them 
\ba
\f{\mathrm{d}\,t}{\mathrm{d}s_\pm}=\{t,C_\pm\}=\mp\f{1}{2 h}\,,\q\q\q\q \f{\mathrm{d}\,q}{\mathrm{d}s_\pm}=\{q,C_\pm\}=1\label{tpm}
\ea
yields exactly the relational observables in (\ref{TX}) restricted to $\cc_\pm\subset\cc$ (and rewritten using (\ref{CH}))
\ba
T_\pm(X):=t(s_\pm)\left|_{q(s_\pm)=X}\right. = t_0\mp \f{X-q_0}{2\sqrt{-p_t}}\,,\q\q\q\q\q P_{T_\pm}(X):=p_t(s_\pm)\left|_{q(s_\pm)=X}\right. =p_t\,.\label{TX2}
\ea

We can thus now separately gauge fix $\cc_\pm$ to construct separate reduced phase spaces for the left and right moving solutions $T_+(X)$ and $T_-(X)$, respectively. We have to accept, of course, that the stationary $p=0$ orbits are ignored so that the result will, again, not be strictly equivalent to the space of orbits $\cp_{\rm red}=\cc/\sim$ (see (a)). But this is as good as it gets for constructing reduced phase spaces that describe the physics relative to the choice of $q$ as a temporal reference system.

Proceeding now in analogy to sec.\ \ref{sec_clttime}, we can gauge fix to, e.g., $q=0$, which is a valid gauge condition for $p\neq0$. The Dirac bracket for any functions $F,G$ on $\cc_\pm$ is again simple
\ba
\{F,G\}_{D_\pm}=\{F,G\}-\{F,C_\pm\}\{q,G\}+\{F,q\}\{C_\pm,G\}\,\label{Dpm}
\ea
and we can consistently remove the now redundant reference system variables $(q,p)$ because
\ba
\{q,p\}_{D_\pm}=\{q,g(t,p_t)\}_{D_\pm}=0\,,\q\q\{p,g(t,p_t)\}_{D_\pm}=\mp\{h(p_t),g(t,p_t)\}_{D_\pm}\,,\q\q \{t,p_t\}_{D_\pm}=\{t,p_t\}=1\,.\nn
\ea
We retain the now evolving $(t,p_t)$ to coordinatize the gauge-fixed two-dimensional reduced phase spaces for the left and right moving solutions, which we shall denote by $\cp_\pm$. Notice that $\cp_\pm\simeq\mathbb{R}\times\mathbb{R}_-$, given that $p_t\leq0$ by (\ref{CH}).

In fact, we need to be slightly more careful. These reduced phase spaces miss their $p_t=0$ boundaries, the way we have constructed them through gauge-fixing. Indeed, on account of (\ref{tpm}, \ref{Dpm}), $\{t,p_t\}_{D_\pm}=1$ {\it only} holds for $p_t<0$ and is undefined for $p_t=0$. However, there is a way to regularize these phase spaces and to add their $p_t=0$ boundaries by switching to a new set of basic phase space variables $(\ft,p_t)$,
where
\ba
\ft:=t\,p_t\,,\q\q\q\Rightarrow\q\q\q\{\ft,p_t\}_{D_\pm}=p_t\,,\q\forall\,p_t\leq0\,.\label{affine}
\ea
Hence, we now have an affine, rather than canonical algebra and it is valid for all $p_t\leq0$. Choosing henceforth $(\ft,p_t)$ as the fundamental degrees of freedom to coordinatize $\cp_\pm$ and their affine algebra to fundamentally define the bracket structure on these phase spaces, we can therefore safely add the $p_t=0$ boundary and consider $\cp_\pm=\mathbb{R}\times\mathbb{R}_-$ in their entirety. In particular, {\it defining} then $t:=\ft/p_t$ yields a derived canonical structure for $(t,p_t)$ on all of $\cp_\pm$, incl.\ its boundary. Henceforth, we shall think of the reduced phase spaces $\cp_\pm$ in this regularized form, being fundamentally defined through an affine algebra, and this will also become crucial in the quantum theory below.

Although $\cp_\pm$ embed into $\cc_\pm$ as a single slice, $\cg_{q=0}\cap\cc_\pm$, we again preserve all the dynamical information. Evaluating the two parameter families (\ref{TX2}) on $\cp_\pm$ can be done by setting $q=0$ and yields\footnote{Note that $T_\pm=t_0\mp \f{q_0}{2\sqrt{-p_t}}= t(s_\pm)\mp \f{q(s_\pm)}{2\sqrt{-p_t}}$ is a constant of motion of $C_\pm$ and so can be evaluated equivalently anywhere on the orbit.}
\ba
T_\pm(X) = t\mp \f{X}{2\sqrt{-p_t}}\,,\q\q\q\q\q P_{T_\pm}(X)=p_t\,.\label{TX3}
\ea
Differentiation with respect to $X$ yields their equations of motion on $\cp_\pm$
\ba
\f{\mathrm{d}T_\pm}{\mathrm{d}X}=\mp \f{1}{2\sqrt{-p_t}}=\{T_\pm,\pm h\}_{D_\pm}\,,\q\q\q\q\q \f{\mathrm{d}P_{T_\pm}}{\mathrm{d}X}=0=\{P_{T_\pm},\pm h\}_{D_\pm}\,,
\ea
which are generated by the {\it physical} Hamiltonian $H_\pm=\pm h=\pm\sqrt{-P_{T_\pm}}=\pm\sqrt{-p_t}$.

In this deparametrized form, the relational dynamics now admits an immediate interpretation as the physics described relative to the `clock' $q$, whose dynamical degrees of freedom, being the reference system, are redundant and have been consistently removed.

\subsection{Switching between relational evolution in $t$ and $q$ time}\label{sec_clswitch}

On the reference-system-neutral constraint surface $\cc$, we thus have the two canonical pairs of relational Dirac observables $(Q(\tau),P(\tau))$ in (\ref{QT}) and $(T(X),P_T(X))$ in (\ref{TX}). The pairs are of course dependent due to the redundancy and we will have to switch between them, when changing from $t$ to $q$ time, or vice versa. Here, we shall explain how to  switch between them at the level of the reduced phase spaces.

To this end, it is necessary to map both pairs into the reduced phase spaces $\cp_{q|t}$ and $\cp_{\pm}$. We find
\be\label{onboth}
\begin{split}
(Q(\tau)=q+2p\tau,\,\,P(\tau)=p)\,,\q\q \left(T(X)=\f{X-q}{2p},\,\,P_T(X)=-p^2\right)\,,\q\q\text{on $\cp_{q|t}$}\,\\
\left(Q_\pm(\tau)=\mp2\sqrt{-p_t}(\tau-t),\,\,P(\tau)=\mp\sqrt{-p_t}\,\right)\,,\q\q\left(T_\pm(X)=t\mp\f{X}{2\sqrt{-p_t}},\,\,P_{T_\pm}(X)=p_t\right)\,,\,\,\text{on $\cp_\pm$}\,.
\end{split}\ee

\begin{figure*}
	\begin{center}
		\includegraphics[scale=0.55]{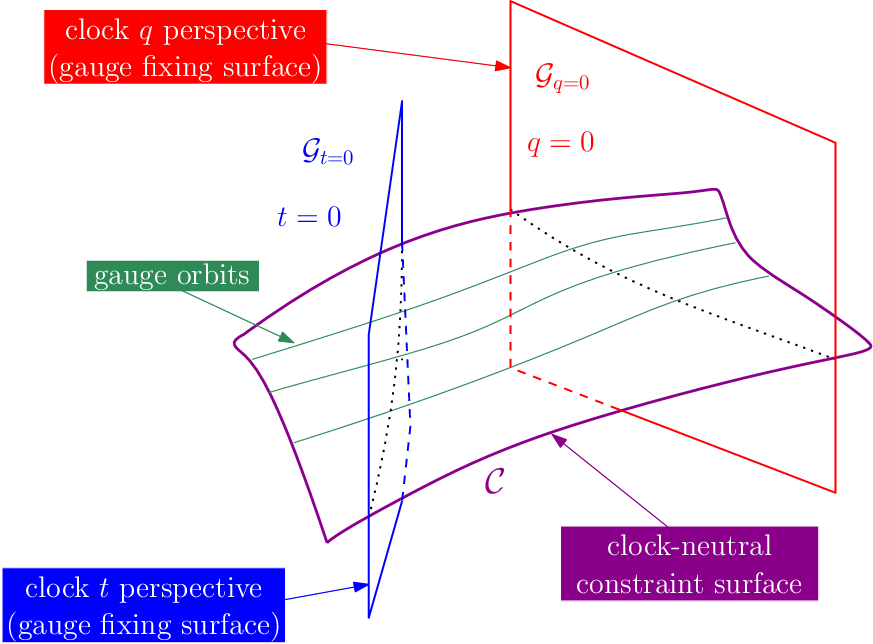}
		\caption{\label{fig_sgeometry}\small{Schematic phase space setup for changing between relational evolutions relative to clocks $t$ and $q$.}}
	\end{center}
\end{figure*}

Suppose we wish to switch from evolution in $t$ to evolution in $q$. Then we also want to switch from $(Q(\tau_f),P(\tau_f))$ on $\cp_{q|t}$, where $\tau_f$ is the final value of $\tau$ up to which we evolve in $t$, to $(T_\pm(X_i),P_{T_\pm}(X_i))$ on $\cp_\pm$, where $X_i$ is the initial value of $X$ with which we continue the evolution in $q$ after the switch. This requires two ingredients: (1) a map that takes a given state $(q,p)$ in $\cp_{q|t}$ to the corresponding state $(t,p_t)$ in $\cp_\pm$, where corresponding means that both states lie on the same gauge orbit in $\cc$ once appropriately embedded into it; (2) a way to determine $X_i$, given $\tau_f$. The situation is summarized in fig.~\ref{fig_sgeometry}.

(1) Constructing the map $\cs_{t\to q\pm}:\cp_{q|t}\rightarrow\cp_\pm$ is straightforward and we detail it in appendix \ref{app_clswitch}. In summary, $\cp_{q|t}$ canonically embeds into $\cc$ as $\cc\cap\cg_{t=0}$; denote the corresponding embedding map by $\iota_{q|t}$. Similarly, $\cp_\pm$ embed canonically as $\cc\cap\cg_{q=0}$ into $\cc$; denote the corresponding embedding maps by $\iota_\pm$. We thus need the gauge transformation $\alpha_{t\to q}:\cc\cap\cg_{t=0}\rightarrow\cc\cap\cg_{q=0}$, generated by $C_H$, to construct the map $\cs_{t\to q\pm}$. Furthermore, we need the projection $\pi_\pm:\cc\supset\cc_\pm\cap\cg_{q=0}\rightarrow\cp_\pm$, which satisfies $\pi_\pm\circ\iota_\pm=\text{Id}_{\cp_\pm}$ and drops all redundant data from $\cc$. Then we find
\ba
\cs_{t\to q\pm}:=\pi_\pm\circ\alpha_{t\to q}\circ\iota_{q|t}\,.
\ea
In appendix \ref{app_clswitch}, we show that in coordinates this map reads
\ba
(q,p)\mapsto (t=-\f{q}{2p},\,p_t=-p^2)\,.\label{clt2q}
\ea
It clearly is a transformation that depends on the relation between the clock and evolving degrees of freedom at the moment of clock change. Notice that $q=0$ intersects of course both $\cc_+\subset\cc$ and $\cc_-\subset\cc$ (modulo the issues for $p=0$) and so $\cs_{t\to q\pm}$ indeed maps the $p<0$ part (the left moving sector) of the phase space $\cp_{q|t}$ to $\cp_+$ and the $p>0$ part (the right moving sector) of the phase space $\cp_{q|t}$ to $\cp_-$. This completes ingredient (1), which can be summarized in
the following commutative diagram:

\begin{center}
\begin{tikzcd}[row sep=huge, column sep = huge]
& \cp_{\rm red}=\mathcal{C}/\sim \arrow[rd, "\zeta_{q=0}"]& \\
\mathcal{C} \cap \mathcal{G}_{t=0} \arrow[ru, "\zeta^{-1}_{t=0}"] \arrow[rr, "\alpha_{t\to  q}"]&& \mathcal{C} \cap \mathcal{G}_{q=0}  \arrow[d, "\pi_{\pm}"]\\
\mathcal{P}_{q|t} \arrow[u, "\iota_{q|t}"] \arrow[rr, "\cs_{t \to q\pm}"] && \mathcal{P}_{\pm}

\end{tikzcd}
\end{center}
Here we have completed the diagram with a `roof' for better comparison with the quantum theory later where the role of the reduced (or physical) phase space $\cp_{\rm red}$ will be taken over by the physical Hilbert space of solutions to the constraint. $\zeta_{t=0}$ is the invertible map which associates to each orbit in $\cc$ its intersection point with the gauge fixing surface $\cg_{t=0}$ (see also \cite{Vytheeswaran:1994np} for a discussion of unfixing gauges in constrained systems). Similarly, $\zeta_{q=0}$ associates to each orbit its intersection point with $\cg_{q=0}$. For $p=0$, this map misses the measure zero set of orbits with $q=const\neq0$ and is thus not defined on the entire phase space. Defining $\varphi_t:=\pi_{t=0}\circ\zeta_{t=0}$, for $\pi_{t=0}$ the projection satisfying $\pi_{t=0}\circ \iota_{q|t}=\text{Id}_{q|t}$, and $\varphi_\pm:=\pi_\pm\circ\zeta_{q=0}$, we thus see that the clock change can also be written as
\ba
\cs_{t\to q\pm} =\varphi_\pm\circ\varphi^{-1}_t\,,
\ea
thereby taking the same compositional form as coordinate changes on a manifold, except that here the role of the manifold is played by the physical phase space $\cp_{\rm red}$, which we can also regard as the clock-neutral phase space. Just like a coordinate map, $\varphi_\pm$ is not defined globally.

(2) The initial value for $X_i$ for the subsequent continued evolution in $X$ on $\cp_\pm$ is simply the image $Q_\pm(\tau_f)=\cs_{t\to q\pm}(Q(\tau_f))$. Indeed, using (\ref{onboth}), we consistently find $T_\pm(X_i=Q_\pm(\tau_f))=\tau_f$, so that the initial value for $t$ in `$q$-time' coincides with the final value of the evolution parameter $\tau$ on $\cp_{q|t}$ prior to the switch. We thus have a continuous relational evolution although switching clocks. This will no longer be the case in the quantum theory later.

Conversely, suppose we wish to switch from $q$ to $t$ time. In complete analogy, the corresponding switch from $(T_\pm(X_f),P_{T_\pm}(X_f))$ on $\cp_\pm$ to $(Q(\tau_i),P(\tau_i))$ on $\cp_{q|t}$, requires the gauge transformation $\alpha_{q\to t}:\cc\cap\cg_{q=0}\rightarrow\cc\cap\cg_{t=0}$ and proceeds via the map (see appendix \ref{app_clswitch} for an explicit construction)
\ba
\cs_{q\pm\to t}:=\pi_{t=0}\circ \alpha_{q\to t}\circ\iota_\pm:\cp_\pm\rightarrow\cp_{q|t}\,,\nn\\
(t,p_t)\mapsto(q=\pm2\,t\,\sqrt{-p_t},\,p=\mp\sqrt{-p_t})\,,\label{clq2t}
\ea
which satisfies the following commutative diagram:
\begin{center}
\begin{tikzcd}[row sep=huge, column sep = huge]
& \cp_{\rm red}=\mathcal{C}/\sim \arrow[rd, "\zeta_{t=0}"]& \\
\mathcal{C} \cap \mathcal{G}_{q=0} \arrow[ru, "\zeta^{-1}_{q=0}"] \arrow[rr, "\alpha_{q\to  t}"]&& \mathcal{C} \cap \mathcal{G}_{t=0}   \arrow[d, "\pi_{t=0}"]\\
\mathcal{P}_{\pm} \arrow[u, "\iota_{\pm}"] \arrow[rr, "\cs_{q\pm \to t}"] && \mathcal{P}_{q|t}
\end{tikzcd}
\end{center}
Using (\ref{onboth}), we also consistently find $Q(\tau_i=T(X_f))=X_f$, where $T(X_f)=\cs_{q\pm\to t}(T_\pm(X_f))$.

In conjunction, this provides a systematic method for consistently switching between the classical relational evolutions in $q$ and $t$ times. Notice that the change from the evolution relative to one temporal reference system to the evolution relative to another always proceeds via the clock-choice-neutral constraint surface $\cc$ or the clock-choice-neutral phase space $\cp_{\rm red}$. This is in harmony with the observation in \cite{Vanrietvelde:2018pgb} that a change of reference frame perspective in relational physics always proceeds via a perspective-neutral structure.

\section{Quantum relational dynamics}\label{sec_q}

We shall now promote all these classical structures, incl.\ the clock switches into the quantum theory. In particular, we begin by quantizing the reduced phase spaces $\cp_{q|t}$ and $\cp_\pm$ and their corresponding relational dynamics using the reduced method. Subsequently, we Dirac quantize the parametrized particle and construct its physical Hilbert space (i.e.\ space of quantum solutions to the constraint), which we shall interpret as the clock-neutral structure in the quantum theory. Thereupon, we demonstrate how to construct the maps from the physical Hilbert space to the clock based Hilbert spaces of the reduced theories through a quantum symmetry reduction procedure, which constitutes a quantum deparametrization. Since these will be invertible (some not globally), we will finally be able to employ these maps to also construct the transformations between the reduced Hilbert spaces associated to different clock choices, completing the quantum clock switches. Just as in the classical case, the quantum clock transformations will proceed via the clock-neutral physical Hilbert space. This result thereby emphasizes that {\it both} the Dirac quantized and quantum deparametrized descriptions are necessary and combine to a complete relational quantum theory that permits switching reference systems, substantiating the claims in \cite{Vanrietvelde:2018pgb, Vanrietvelde:2018dit}. 

As emphasized in the introduction, the present model is a special case where Dirac and reduced quantization are actually equivalent, with the equivalence maps established through the quantum symmetry reduction procedure. In more general models, Dirac and reduced quantization are not equivalent \cite{Ashtekar:1982wv,Ashtekar:1991hf,Schleich:1990gd,Kunstatter:1991ds,Hajicek:1990eu,Romano:1989zb,Dittrich:2016hvj,Dittrich:2015vfa,Loll:1990rx,Kaminski:2009qb,Domagala:2010bm,Giesel:2016gxq}. However, this is not a problem for the present method of clock changes: in general, our proposal for the procedure of changing quantum reference systems is to start with Dirac quantization and to always use the quantum symmetry reduction maps as the `quantum coordinate maps' taking the clock-neutral description of the physical Hilbert space into the description of the quantum dynamics relative to the chosen clock. We give primacy to Dirac quantization because it is more general in quantizing all degrees of freedom \emph{a priori}, i.e.\ also including whatever one may choose as the temporal reference system in the end. In particular, as we will see shortly, it will provide us with the same redundancy on the `quantum constraint surface' (i.e.\ the physical Hilbert space) that we had classically on the constraint surface $\cc$ and this once more will give us the ability to describe the same physics in many different ways; redundancy in the description is the prerequisite for quantum clock covariance. 

In the literature it is sometimes argued that Dirac quantization is more general (and thus favourable) because in the reduced method quantum fluctuations of the reference system (in our case the clock) are impossible to begin with given that its dynamical degrees of freedom are eliminated prior to quantization. However, this argument is somewhat misleading and this once more has to do with redundancy: while it is true that in Dirac quantization the reference degrees of freedom (and their fluctuations) are independent on the total (kinematical) Hilbert space, on the physical Hilbert space not all degrees of freedom and thus not all fluctuations are independent. Specifically, when we choose the clock as our temporal reference system, its degrees of freedom and fluctuations will be the dependent ones, determined through the evolving degrees of freedom. In other words, upon solving the constraints, in Dirac quantization too the reference system will not feature independent quantum fluctuations. This is also the reason why in case of equivalence (as in the model of this article) we can recover the reduced quantum theory from the Dirac quantum theory without discarding independent information. The rationale for giving Dirac quantization primacy over reduced quantization can thus not be based on fluctuations of the reference system. Instead, in our opinion, the genuine reason for giving primacy to Dirac quantization is the redundancy it induces in the quantum theory, thereby providing the ground for a quantum covariance of reference systems.

The reason for the typical difference between Dirac and reduced quantization is rather rooted in the fact that imposing restrictions (in this case coming from the constraint(s)) on the permissible values of observables can lead to quite different results depending on whether this is done before or after quantization as this interplays subtly with boundary conditions on wave functions. It is only a coincidence that in simple models, such as in the present manuscript and in \cite{Vanrietvelde:2018pgb,pacosmo}, the quantum symmetry reduced theories coincide with the quantization of the classically symmetry reduced theories. This point is further discussed in \cite{Vanrietvelde:2018pgb, Vanrietvelde:2018dit,Hoehn:2019owq,Hoehn:2020epv}.

\begin{figure}
\centering
    \begin{tikzcd}[row sep=large, column sep = large]
& & \textrm{original phase space } \mathbb{R}^4 \arrow{lld}[swap]{C_+ = q =0} \arrow[ld, "C_- =q =0"] \arrow[d,"\textrm{Dirac quantization}"] \arrow[rrd, "C_H = t =0"] & & \\
\mathcal{P}_+ \arrow[d,"\textrm{\hspace*{.43cm}affine reduced quantization}"] & \mathcal{P}_- \arrow{d} & \mathcal{H}_\textrm{kin} \arrow[dd,"\delta(\hat{C}_H)"] & & \mathcal{P}_{q|t} \arrow[dd,"\textrm{\hspace*{-2.3cm}canonical reduced  quantization}"] \\
\mathcal{H}_+ & \mathcal{H}_- & & & \\
& \mathcal{H}_\textrm{phys}^{t|q} \arrow[lu, "_q\bra{q=0} \theta(- \hat{p}) (- \hat{p}_t)^\frac{1}{4}"] \arrow{u}[swap]{_q\bra{q=0} \theta( \hat{p}) (- \hat{p}_t)^\frac{1}{4}} & \mathcal{H}_\textrm{phys} \arrow[l, "\mathcal{T}_q"] \arrow{r}[swap]{\mathcal{T}_t} & \mathcal{H}_\textrm{phys}^{q|t} \arrow{r}[swap]{{}_t\bra{t=0}} & \mathcal{H}_{q|t}
\end{tikzcd}
\caption{{\small Overview of the various steps of the Dirac quantization, the three reduced quantizations, as well as their links. All steps are explained in detail in the main text. $\cp_\pm$ are the right/left mover reduced phase spaces and $\ch_\pm$ their quantizations, while $\cp_{q|t}$ and $\ch_{q|t}$ are the reduced phase and Hilbert space relative to clock $t$, respectively. $\ch_{\rm kin}$ and $\ch_{\rm phys}$ are the kinematical and physical Hilbert space of the Dirac quantization, respectively. Mapping from $\ch_{\rm phys}$ to the reduced Hilbert spaces involves a trivialization $\ct_t$ or $ \ct_q$ of the constraint to the chosen clock and a subsequent projection onto the classical gauge fixing conditions. }} \label{fig1}
\end{figure}
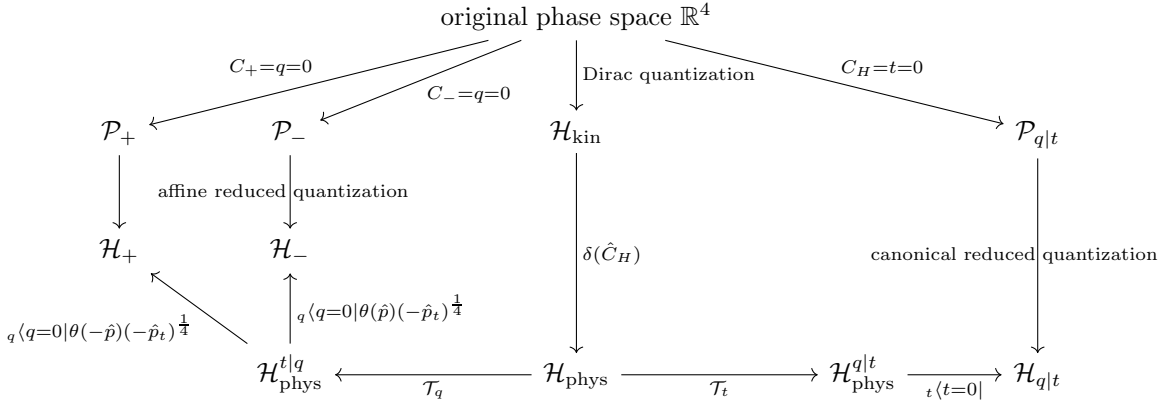

Along the way of the construction, we will encounter a number of transformations, projections and Hilbert spaces. For better orientation and visualization of the following procedure, we organize the various ingredients and their relation in fig.\ \ref{fig1}.

\subsection{Reduced quantization in $t$ time}\label{sec_qttime}

It is standard to quantize the reduced phase space $\cp_{q|t}$ of the parametrized free particle in sec.\ \ref{sec_clttime}, given that it coincides with the phase space of the usual unparametrized Newtonian particle. We promote the Dirac bracket $\{.,.\}_D$ to a commutator $[.,.]$ and $(q,p)$ to conjugate operators $[\hat{q},\hat{p}]=i$\footnote{Henceforth, we work in units where $\hbar=1$.} on a Hilbert space $\ch_{q|t}:=L^2(\mathbb{R})$. To later link with the Dirac quantized theory, it will be more convenient to choose the momentum representation, in which states take the form
\ba
\ket{\psi}_{q|t}=\int\,\mathrm{d}p\,\psi_{q|t}(p)\,\ket{p}\,\label{redtstate}
\ea
and the inner product reads
\ba
\la\phi|\psi\ra_{q|t}=\int\,\mathrm{d}p\,\phi^*_{q|t}(p)\,\psi_{q|t}(p)\,.\label{redtip}
\ea
Quantizing the evolving constants of motion (\ref{QT2}) produces the evolving operators 
\ba
\hat{Q}(\tau)=\hat{q}+2\hat{p}\tau\,,\q\q\q\q\q\q\hat{P}(\tau)=\hat{p}\,,\label{redqQT}
\ea
in the Heisenberg picture, satisfying the Heisenberg equations with Hamiltonian $\hat{H}=\hat{P}^2(\tau)=\hat{p}^2$,
\ba
\f{\mathrm{d}\hat{Q}}{\mathrm{d}\tau}=-i\,[\hat{Q},\hat{H}]=2\hat{P}\,,\q\q\q\q\f{\mathrm{d}\hat{P}}{\mathrm{d}\tau}=-i\,[\hat{P},\hat{H}]=0\,.
\ea
In the Schr\"odinger picture, states will satisfy the corresponding Schr\"odinger equation.

\subsection{Reduced quantization in $q$ time}\label{sec_qqtime}

Quantizing the reduced phase spaces relative to $q$ is more complicated, since $\cp_\pm\simeq\mathbb{R}\times\mathbb{R}_-$. In particular, $\hat{t}$ cannot be promoted to a self-adjoint operator that is conjugate to $\hat{p}_t$ since otherwise it would map states with support on $p_t<0$ to states with support on the classically forbidden region $p_t>0$ \cite{isham2}. Instead, we recall from sec.\ \ref{sec_clqtime} that the regularized $\cp_\pm$ is fundamentally defined through an affine algebra. We will thus resort to {\it affine} quantization \cite{isham2}. 

Our aim is therefore to promote the Dirac brackets $\{.,.\}_{D_\pm}$ to a commutator and $\ft,p_t$ in (\ref{affine}) to operators that satisfy $[\hat{\ft},\hat{p}_t]=i\,\hat{p}_t$. Again, it will be more convenient to work in momentum representation, in which we can employ $\ch_{\pm}=L^2(\mathbb{R}_-,\f{\mathrm{d}p_t}{-p_t})$ as a Hilbert space and represent states as follows
\ba
\ket{\psi}_\pm=\int_{-\infty}^0\,\f{\mathrm{d}p_t}{-p_t}\,\psi_\pm(p_t)\,\ket{p_t}_\pm\,,\label{affstate}
\ea
with a scale-invariant measure that carries a minus sign so it is positive. Notice that the generalized momentum eigenstates are here normalized as
\ba
{}_\pm\la p_t|p_t'\ra_\pm=-p_t\,\delta(p_t-p_t')\,,\label{unnorm}
\ea 
so that the inner product reads
\ba
\la\phi|\psi\ra_\pm=\int_{-\infty}^0\,\f{\mathrm{d}p_t}{-p_t}\,\phi^*_\pm(p_t)\,\psi_\pm(p_t)\,,\label{affpip}
\ea
and 
\ba
\mathds{1}=\int_{-\infty}^0\,\f{\mathrm{d}p_t}{-p_t}\,\ket{p_t}_\pm{}_\pm\bra{p_t}\,,\q\q\q\Rightarrow\q\q\q \mathds{1}\,\ket{p'_t}_\pm=\int_{-\infty}^0\,{\mathrm{d}p_t}\,\delta(p_t-p_t')\ket{p_t}_\pm=\ket{p_t'}_\pm\,.
\ea
On this Hilbert space we can represent our basic variables as
\ba
\hat{\ft}\,\psi_\pm(p_t)=i\,p_t\,\f{\p}{\p\,p_t}\,\psi_\pm(p_t)\,,\q\q\q\q \hat{p}_t\,\psi_\pm(p_t)=p_t\,\psi_\pm(p_t)\,,\q\q\q\Rightarrow\q\q[\hat{\ft},\hat{p}_t]=i\,\hat{p}_t\,,\label{affrep}
\ea
and in this form $\hat{\ft}$ and $\hat{p}_t$ are self-adjoint \cite{isham2}.

Next, we wish to promote the reduced evolving constants of motion (\ref{TX3}) to evolving operators in the Heisenberg picture on $\ch_\pm$. Here, we have to be careful, as $\hat{t}$ can not be a self-adjoint operator and we also have to quantize $(-p_t)^{-1/2}$. A na\"ive quantization $\widehat{(\sqrt{-p_t})^{-1}}$ via spectral decomposition will neither yield a self-adjoint operator as it becomes unbounded for $p_t\rightarrow0$. All these pathologies, of course, have their origin in the $p_t\rightarrow0$ limit, which already classically caused trouble. Indeed, it follows from the discussion in \cite{Grot:1996xu,aharonov1998measurement,muga2000arrival,Gambini:2000ht } that a na\"ive quantization of time-of-arrival functions produces operators, which are neither self-adjoint, nor possess self-adjoint extensions. We therefore have to regularize these operators carefully to obtain a well-defined and self-adjoint quantum version of $T_\pm(X)$. In that case we can interpret $\hat{T}_\pm(X)$ as a genuine quantum observable with a consistent probabilistic interpretation of expectation values 
and a spectral decomposition.\footnote{Alternatively, one could also try to develop a description in terms of positive operator-valued measures.} 

When regularizing these operators, we wish to do so in a minimal manner by modifying them only in an infinitesimal neighbourhood of the troublesome boundary $p_t=0$, such that the regularized $\hat{t}$ will be arbitrarily close to being canonically conjugate to $\hat{p}_t$ and such that the regularized $\hat{T}_\pm(X)$ will be arbitrarily close to the reduced observables $T_\pm(X)$ in the classical limit.

First, classically we had $t:=\ft/p_t$, so we now quantize and regularize it as follows on $\ch_\pm$:\footnote{For later convenience, we attach a label $\pm$ to this operator although it is here not strictly necessary. However, later it will help to distinguish it from a similarly defined operator in the Dirac quantized theory.}
\ba
\hat{t}_{\delta\pm}:=\f{1}{2}\left(\widehat{(p_t)_\delta^{-1}}\,\hat{\ft}+\hat{\ft}\,\widehat{(p_t)_\delta^{-1}}\right)\,,\label{regtime0}
\ea
where $\delta>0$ is an arbitrarily small positive number and 
\ba\label{invpt}
\widehat{(p_t)_\delta^{-1}}\ket{p_t}_\pm:=\begin{cases}
 \,\,\,\,\,\f{1}{p_t}\ket{p_t}_\pm     & p_t\leq-\delta^2 , \\
 -  \f{1}{\delta^2}\ket{p_t}_\pm   & -\delta^2<p_t\leq0.
\end{cases}
\ea
Since $\widehat{(p_t)_\delta^{-1}}$ by construction has a complete orthogonal basis of generalized eigenstates with real eigenvalues, it is self-adjoint. As a consequence, given that $\hat{t}_{\delta\pm}$ is a symmetrization of two self-adjoint operators, it is self-adjoint too. In particular, we now have
\ba
[\hat{t}_{\delta\pm},\hat{p}_t]=i\, \widehat{(p_t)_\delta^{-1}}\,\hat{p}_t\,,\label{almostcan}
\ea
so that $\hat{t}_{\delta\pm}$ and $\hat{p}_t$ indeed are not exactly canonically conjugate, but arbitrarily close to being canonically conjugate, with modifications only for $-\delta^2<p_t\leq0$.

Second, we quantize and regularize the inverse square root appearing in (\ref{TX3}) as follows on $\ch_\pm$:
\ba\label{invroot}
\widehat{(\sqrt{-p_t})_\delta^{-1}}\ket{p_t}_\pm:=\begin{cases}
\f{1}{\sqrt{-p_t}}\ket{p_t}_\pm     & p_t\leq-\delta^2 , \\
   \f{\sqrt{-p_t}}{\delta^2}\ket{p_t}_\pm   & -\delta^2<p_t\leq0.
\end{cases}
\ea
Again, this operator is self-adjoint. We thus see that $\widehat{(\sqrt{-p_t})_\delta^{-1}}\cdot \widehat{(\sqrt{-p_t})_\delta^{-1}}\neq \widehat{(p_t)_\delta^{-1}}$ when $-\delta^2<p_t\leq0$. However, this is not a problem as it affects only an infinitesimal region and we shall explain shortly why the regularization in this form is needed for dynamical consistency.

We are now in the position to promote the reduced evolving constants of motion (\ref{TX3}) to self-adjoint operators on $\ch_\pm$ in the Heisenberg picture:
\ba
\hat{T}_\pm(X):=\hat{t}_{\delta\pm}\mp\f{X}{2}\,\widehat{(\sqrt{-p_t})_\delta^{-1}}\,,\q\q\q\q\hat{P}_{T_\pm}(X)=\hat{p}_t\,.\label{redqTX}
\ea
They satisfy the following Heisenberg equations with Hamiltonian $\hat{H}_\pm=\pm\hat{h}=\pm\widehat{\sqrt{-P_{T_\pm}}}=\pm\widehat{\sqrt{-p_t}}$
\ba
\f{\mathrm{d}\hat{T}_\pm}{\mathrm{d}X}=\mp\f{1}{2}\,\widehat{(\sqrt{-p_t})_\delta^{-1}}=-i\,[\hat{T}_\pm,\hat{H}_\pm]\,,\q\q\q\q\f{\mathrm{d}\hat{P}_{T_\pm}}{\mathrm{d}X}=0=-i\,[\hat{P}_{T_\pm},\hat{H}_\pm]\,.\label{HeisenX}
\ea
For the right equation this is immediate, for the left it is non-trivial due to the commutator and we show it in appendix \ref{app_commut}. This is where the regularization of the inverse square root in the form (\ref{invroot}) becomes crucial. The reason why we have chosen it in that form is so we have consistent commutator-generated operator evolution equations.
In the Schr\"odinger picture, dynamical states in $\ch_\pm$ will obviously have to satisfy the Schr\"odinger equation corresponding to $\hat{H}_\pm$.

We could, of course, have chosen a different regularization of our operators altogether but the one above will turn out to be convenient for linking with the Dirac quantized theory of the next section.

\subsection{Dirac quantization -- the clock-neutral quantum theory}\label{sec_Dirac}

We shall now first quantize the full classical phase space $\mathbb{R}^4$ of sec.\ \ref{sec_cl} (i.e., the extended phase space of appendix \ref{app_nrpart}), incl.\ unphysical and gauge-dependent degrees of freedom, thereby promote the Hamiltonian constraint to a quantum operator and subsequently solve it in the quantum theory. Just like $\cc$ is the clock-choice-neutral structure of the classical theory, the result will yield the clock-choice-neutral quantum structures via which we will later switch temporal reference systems.

We thus promote $(q,p)$ and $(t,p_t)$ to conjugate operators $[\hat{q},\hat{p}]=[\hat{t},\hat{p}_t]=i$ on a kinematical Hilbert space $\ch_{\rm kin}:=L^2(\mathbb{R}^2)$. The Hamiltonian constraint (\ref{CH}) thus becomes an operator and we require that {\it physical} states are characterized by solving it in quantum form
\ba
\hat{C}_H\,\ket{\psi}_{\rm phys}=(\hat{p}_t+\hat{p}^2)\,\ket{\psi}_{\rm phys}\overset{!}{=}0\,.\label{qCH}
\ea
Such states are then immediately reparametrization-invariant since $\exp(i\,s\,\hat{C}_H)\,\ket{\psi}_{\rm phys}=\ket{\psi}_{\rm phys}$. Hence, while classically solving the constraint leads to the constraint surface $\cc$ on which we still have the gauge flows generated by $C_H$, we see that in the quantum theory solving the constraint automatically leads to gauge-invariance. This is, of course, a consequence of the uncertainty relations: $\hat{C}_H$ has a continuous spectrum so that its zero-eigenstates will be maximally spread over gauge degrees of freedom, which are conjugate to it. 

Owing to the continuity of $\hat{C}_H$'s spectrum, physical states will {\it not} be normalized in the inner product on $\ch_{\rm kin}$. We thus have to construct a new one for the space of solutions to (\ref{qCH}). To this end, we resort to {\it group averaging} (or refined algebraic quantization) \cite{Marolf:1995cn,Marolf:2000iq,Thiemann:2007zz}\footnote{See also \cite{Kempf:2000qz} for an alternative method.}, defining the (improper) projector
\ba
\delta(\hat{C}_H)=\f{1}{2\pi}\,\int_{-\infty}^{\infty}\,\mathrm{d}s\,\exp(i\,s\,\hat{C}_H)\,,\q\q\q\q \delta(\hat{C}_H):\ch_{\rm kin}\rightarrow\ch_{\rm phys}\,,
\ea
onto the {\it physical} Hilbert space $\ch_{\rm phys}$, which will be constructed out of the solutions to (\ref{qCH}).

In momentum representation, this yields explicitly
\ba
\ket{\psi}_{\rm phys}&=&\delta(\hat{C}_H)\,\ket{\psi}_{\rm kin}=\f{1}{2\pi}\,\int_{-\infty}^{\infty}\,\mathrm{d}s\,\exp(i\,s\,\hat{C}_H)\,\int\,\mathrm{d}p\,\mathrm{d}p_t\,\psi_{\rm kin}(p,p_t)\,\ket{p}_q\ket{p_t}_t\nn\\
&=&\int\,\mathrm{d}p\,\mathrm{d}p_t\,\delta(p_t+p^2)\,\psi_{\rm kin}(p,p_t)\,\ket{p}_q\ket{p_t}_t\,.\label{proj}
\ea
There is thus a redundancy in the representation of physical states and for linking with the reduced quantum theories in $t$ and $q$ times later, we will solve it in two ways. Firstly, we can write
\ba
\ket{\psi}_{\rm phys}=\int_{-\infty}^{\infty}\,\mathrm{d}p\,\psi_{\rm kin}(p,-p^2)\,\ket{p}_q\ket{-p^2}_t\,.\label{phys1}
\ea
However, using (\ref{factorize}) and
\ba
\delta(p_t+p^2)=\f{\delta(C_+)}{2\sqrt{-p_t}}+\f{\delta(C_-)}{2\sqrt{-p_t}}\,,
\ea
we can write the same physical state equivalently as 
\ba
\ket{\psi}_{\rm phys}=\int_{-\infty}^0\,\f{\mathrm{d}p_t}{2\sqrt{-p_t}}\,\Big(\psi_{\rm kin}(-\sqrt{-p_t},p_t)\ket{-\sqrt{-p_t}\,}_q\,\ket{p_t}_t+\psi_{\rm kin}(\sqrt{-p_t},p_t)\,\ket{\sqrt{-p_t}\,}_q\ket{p_t}_t\Big)\,.\label{phys2}
\ea

The {\it physical} inner product on the space of solutions to (\ref{qCH}) is defined via \cite{Marolf:1995cn,Marolf:2000iq,Thiemann:2007zz}
\ba
(\phi_{\rm phys},\psi_{\rm phys})_{\rm phys}:={}_{\rm kin}\la\phi|\,\delta(\hat{C}_H)\,|\psi\ra_{\rm kin}\,,
\ea
where ${}_{\rm kin}\la\cdot|\cdot\ra_{\rm kin}$ denotes the standard $L^2$ inner product on $\ch_{\rm kin}$. Indeed, thanks to the symmetry of $\delta(\hat{C}_H)$, this inner product is well-defined on equivalence classes of states in $\ch_{\rm kin}$, where a given equivalence class contains all the kinematical states that are projected via (\ref{proj}) to the same physical state. Upon Cauchy completion (plus dividing out spurious solutions and zero-norm states) one can thereby turn the space of solutions to (\ref{qCH}) to a genuine Hilbert space $\ch_{\rm phys}$ \cite{Marolf:1995cn,Marolf:2000iq,Thiemann:2007zz}.

In particular, in our two representations, we can equivalently write
\ba\label{PIP}
(\phi_{\rm phys},\psi_{\rm phys})_{\rm phys}&=&\int_{-\infty}^{\infty}\,\mathrm{d}p\,\phi^*_{\rm kin}(p,-p^2)\,\psi_{\rm kin}(p,-p^2)\\
&=&\int_{-\infty}^0\,\f{\mathrm{d}p_t}{2\sqrt{-p_t}}\,\Big[\phi^*_{\rm kin}(-\sqrt{-p_t},p_t)\,\psi_{\rm kin}(-\sqrt{-p_t},p_t)+\phi^*_{\rm kin}(\sqrt{-p_t},p_t)\,\psi_{\rm kin}(\sqrt{-p_t},p_t)\Big]\,.\nn
\ea
That is, in the latter $p_t$ based representation, we get a separate inner product for the left and right moving modes.

Lastly, we need to worry about observables on the physical Hilbert space $\ch_{\rm phys}$. Notice that an observable $\hat{O}$ on $\ch_{\rm phys}$ has to be gauge invariant $[\hat{O},\hat{C}_H]=0$, for otherwise its action would map out of the zero-eigenspace of the constraint $\hat C_H$, i.e.\ $\ch_{\rm phys}$. Hence, we need to work with quantum Dirac observables. In particular, we would like to represent the two classical families of relational Dirac observables (\ref{QT}, \ref{TX}) as families of quantum Dirac observables on $\ch_{\rm phys}$. 
For (\ref{QT}) this is simple:
\ba
\hat{Q}(\tau)=2\,\hat{p}\,(\tau-\hat{t})+\hat{q}\,,\q\q\q\q\q\hat{P}(\tau)=\hat{p}\,,\label{qQT}
\ea
with $[\hat{Q}(\tau),\hat{C}_H]=[\hat{P}(\tau),\hat{C}_H]=0$, directly producing self-adjoint operators on $\ch_{\rm phys}$.

For (\ref{TX}) this is more involved because of factor ordering and the inverse power of $p$. We have to worry about $p=0$ because $\hat{p}$ is a quantum Dirac observable and so $\ch_{\rm phys}$ will contain states with support on it.
Again, we need to resort to a careful regularization of these operators, by only slightly modifying their behavior in an infinitesimal neighbourhood of the troublesome $p=0$ such that (i) in the classical limit they will be arbitrarily close to (\ref{TX}), and (ii) they will later map correctly to the regularized reduced evolving observables of the reduced quantum theory in $q$ time of sec.\ \ref{sec_qqtime}. 

First, we regularize and quantize the inverse momentum (as in \cite{Grot:1996xu}) as follows
\ba\label{invp}
\widehat{(p)_\delta^{-1}}\ket{p}_q:=\begin{cases}
\,\,\f{1}{p}\ket{p}_q     & |p|\geq\delta , \\
   \f{p}{\delta^2}\ket{p}_q   & |p|<\delta\,,
\end{cases}
\ea
where $\delta>0$ is the same arbitrarily small positive number, which we already used in the regularizations of inverse powers of $p_t$ in (\ref{invpt}, \ref{invroot}). It is clear that $\widehat{(p)_\delta^{-1}}$ thus defined, has a complete orthogonal basis of generalized eigenstates $\ket{p}_q\ket{-p^2}_t$ with real eigenvalues on $\ch_{\rm phys}$ and so is self-adjoint.

We could therefore now try to quantize $T(X)$ in (\ref{TX}) through symmetrization as follows
\ba
\hat{T}(X):=\hat{t}+\f{1}{4}\left(\widehat{(p)_\delta^{-1}}\,(X-\hat{q})+(X-\hat{q})\,  \widehat{(p)_\delta^{-1}}\right)\,.
\ea
However, as one can easily check, this would fail to define a quantum Dirac observable because
\ba
[\,\hat{T}(X),\hat{C}_H\,]=i\,(1-\hat{p}\,\widehat{(p)_\delta^{-1}})\,,
\ea
which fails to vanish for $|p|<\delta$ due to the regularization. It is clear that we also have to regularize $\hat{t}$ because its action can map physical states, which do {\it not} have support on $p_t>0$ to states that do, being conjugate to the quantum Dirac observable $\hat{p}_t$ (see the analogous discussion in the reduced theory of sec.\ \ref{sec_qqtime}). On $\ch_{\rm kin}$, we thus define the regularized inverse $p_t$ in complete analogy to (\ref{invpt}) via 
\ba\label{invpt2}
\widehat{(p_t)_\delta^{-1}}\ket{p_t}:=\begin{cases}
 \,\,\,\,\,\f{1}{p_t}\ket{p_t}     & p_t\leq-\delta^2 , \\
 -  \f{1}{\delta^2}\ket{p_t}   & -\delta^2<p_t\leq0\,,
\end{cases}
\ea
which similarly is self-adjoint on $\ch_{\rm phys}$, to then define a regularized `time operator' on $\ch_{\rm kin}$ by
\ba
\hat{t}_\delta:=\f{1}{2}\left(\hat{t}\,\hat{p}_t\,\widehat{(p_t)_\delta^{-1}}+\hat{p}_t\,\widehat{(p_t)_\delta^{-1}}\,\hat{t}\right)\,.\label{regtime}
\ea
In analogy to (\ref{almostcan}), we then have
\ba
[\,\hat{t}_\delta,\hat{p}_t\,]=i\,\hat{p}_t\,\widehat{(p_t)_\delta^{-1}}\,,\label{almostcan2}
\ea
so that the regularized $\hat{t}_\delta$ is `almost' canonically conjugate to the Dirac observable $\hat{p}_t$. This finally permits us to regularize and quantize the evolving constants of motion (\ref{TX}) in the form
\ba
\hat{T}_\delta(X):=\hat{t}_\delta+\f{1}{4}\left(\widehat{(p)_\delta^{-1}}\,(X-\hat{q})+(X-\hat{q})\,  \widehat{(p)_\delta^{-1}}\right)\,,\q\q\q\q\hat{P}_T(X):=\hat{p}_t\,.\label{qTX}
\ea
Using (\ref{invp}, \ref{invpt2}), it is now straightforward to check that 
\ba
[\,\hat{T}_\delta(X),\hat{C}_H\,]=i\,(\hat{p}_t\,\widehat{(p_t)_\delta^{-1}}-\hat{p}\,\widehat{(p)_\delta^{-1}})=0\q\q\text{on $\ch_{\rm phys}$}\,,\q\q\q[\,\hat{P}_T(X),\hat{C}_H\,]=0\,,
\ea
so that (\ref{qTX}) constitute two genuine families of relational quantum Dirac observables.

We propose to regard $\ch_{\rm phys}$ as the clock-choice-neutral quantum structure of the model. Indeed, we now have two sets of relational quantum observables (\ref{qQT}, \ref{qTX}) on $\ch_{\rm phys}$. We thus have a redundancy of observables for describing the system, as well as a redundancy in the representation of physical states and the physical inner product. In fact, we could have constructed other families of relational observables and explicit representations of physical states and the inner product too, had we chosen even more different clock variables. $\ch_{\rm phys}$ thereby encodes a multitude of clock choices at once and is thus not `timeless' as often stated (it is background-timeless, but not internal-timeless). As the clock-neutral Hilbert space, $\ch_{\rm phys}$ provides a global description of the physics, prior to having chosen a temporal reference system relative to which we describe the quantum dynamics of the remaining degrees of freedom.

\subsection{From Dirac to reduced quantum theory in $t$ time}\label{sec_Dirac2t}

Our aim is now to recover the reduced quantum theory of sec.\ \ref{sec_qttime} with its time evolution relative to the clock $t$ from the clock-choice-neutral Dirac quantized theory of the previous section. Recall from sec.\ \ref{sec_clttime} that mapping from the clock-choice-neutral constraint surface $\cc$ to the reduced phase space $\cp_{q|t}$ in $t$ time involved a gauge choice $t=0$ to remove the redundant clock degrees of freedom. We would thus like to emulate this step at the quantum level and remove the degrees of freedom associated to $t$. However, it is already clear that we have to proceed somewhat differently, owing to the observation in sec.\ \ref{sec_Dirac} that $\ch_{\rm phys}$ is already reparametrization-invariant so that we can no longer fix any gauges in the Dirac quantized theory.

As exhibited in \cite{Vanrietvelde:2018pgb, Vanrietvelde:2018dit}, the quantum analog of the classical reduction by gauge fixing is:
\begin{enumerate}
\item {\it Trivialize} the quantum constraint(s) to the reference system. That is, transform them in such a way that they only act on the degrees of freedom of the chosen reference system, which one now considers as redundant.
\item {\it `Project'} onto the classical gauge-fixing conditions. We write projection in quotation marks as it is only a projection (and non-invertible) when applied to the kinematical Hilbert space, but not when applied to the physical Hilbert space. In the latter case, it only removes redundant degrees of freedom which have already been fixed through the constraint.\footnote{Returning to the discussion about the relation between Dirac and reduced quantization at the beginning of this section, note that fluctuations of the reference system are no longer independent, but redundant upon solving the constraint, which is why we can remove them.}
 The `projection' will thereby not remove any independent physical information and can thus also be inverted.
\end{enumerate}
In conjunction, this constitutes a quantum symmetry reduction relative to the chosen reference system, which here amounts to a quantum deparametrization of the model. We shall now illustrate this procedure for clock $t$.

Define the {\it trivialization unitary} 
\ba
\ct_t:=\exp(i\,\hat{t}\,\hat{p}^2)=\exp(i\,\hat{t}\,\hat{H})\,.\label{trivialt}
\ea
This is, of course, essentially the time evolution operator, except that we now have an operator $\hat{t}$ appearing in it. Intuitively, the exponent can be viewed as $-i\,\hat{t}\cdot\hat{p}_t$, except that $\hat{p}_t$ has been replaced by solving the constraint equation (\ref{qCH}) for it in terms of $\hat{p}^2$ so that $\ct_t$ is unitary on $\ch_{\rm kin}$. $\ct_t$ is, however, not unitary on $\ch_{\rm phys}$ as it does not commute with $\hat{C}_H$. Instead, it will define an isometry that maps $\ch_{\rm phys}$ to a new Hilbert space, i.e.\ to a new representation of physical states.

The key property of this map is that it {\it trivializes} the Hamiltonian constraint to the clock $t$:
\ba
\ct_t\,\hat{C}_H\,\ct^\dag_t=\hat{p}_t\,,
\ea
where $\dag$ is defined with respect to $\ch_{\rm kin}$. Correspondingly, using the representation (\ref{phys1}) of physical states, we find
\ba
\ct_t\,\ket{\psi}_{\rm phys}=\int_{-\infty}^{\infty}\,\mathrm{d}p\,\psi_{\rm kin}(p,-p^2)\,\ket{p}_q\ket{0}_t\,.
\ea
The clock factor of the state contains thereby no more relevant information about the original $\ket{\psi}_{\rm phys}$ and has become entirely redundant. We can thus remove it without losing information by `projecting' onto the classical gauge fixing condition $t=0$, in analogy to the Page-Wootters construction \cite{Page:1983uc}\footnote{Equivalence of the present procedure with the Page-Wootters formalism has later been established in \cite{Hoehn:2019owq,Hoehn:2020epv}.}
\ba
\ket{\psi}_{q|t}&:=&\sqrt{2\pi}\,{}_t\bra{t=0}\,\ct_t\,\ket{\psi}_{\rm phys}=\,\int\,\mathrm{d}p_t\,{}_t\bra{p_t}\,\int_{-\infty}^{\infty}\,\mathrm{d}p\,\psi_{\rm kin}(p,-p^2)\,\ket{p}_q\ket{0}_t\nn\\
&=&\int_{-\infty}^{\infty}\,\mathrm{d}p\,\psi_{\rm kin}(p,-p^2)\,\ket{p}_q\,.\label{PaW}
\ea
Upon identifying 
\ba
\psi_{q|t}(p):=\psi_{\rm kin}(p,-p^2)\label{identify}
\ea
we therefore recover the states (\ref{redtstate}) as initial $t=0$ states, or states in the Heisenberg picture.

In agreement with this, we find that the relational quantum observables (\ref{qQT}) transform correctly
\ba
\ct_t\,\hat{Q}(\tau)\,\ct^\dag_t &=& \ct_t\,\left(2\,\hat{p}\,(\tau-\hat{t}\,)+\hat{q}\right)\,\ct^\dag_t=2\,\hat{p}\,\tau+\hat{q}\,,\nn\\
\ct_t\,\hat{P}(\tau)\,\ct^\dag_t &=& \ct_t\,\hat{p}\,\ct^\dag_t=\hat{p}\,
\ea
to their reduced form (\ref{redqQT}) on $\ch_{q|t}$ -- likewise in the Heisenberg picture. Indeed, these transformed observables are compatible with the `projection' onto $t=0$ above.

Finally, as one can easily check, $\ct_t$ and the ensuing projection also preserve the inner product, since
\ba
(\phi_{\rm phys},\psi_{\rm phys})_{\rm phys}&=&{}_{\rm kin}\la\phi|\,\psi\ra_{\rm phys}={}_{\rm kin}\la\ct_t\,\phi|\,\ct_t\,\psi\ra_{\rm phys}={}_{q|t}\bra{\phi}\psi\ra_{q|t}\nn\\
&=&\int_{-\infty}^{\infty}\,\mathrm{d}p\,\phi^*_{\rm kin}(p,-p^2)\,\psi_{\rm kin}(p,-p^2)\,,
\ea
recovering the inner product (\ref{redtip}) on $\ch_{q|t}$ upon the identification (\ref{identify}). The total quantum symmetry reduction map relative to clock $\hat t$ is thus 
\ba
\Phi_t:=\sqrt{2\pi}\,{}_t\bra{t=0}\,\ct_t\,.\label{Phit}
\ea
For later convenience, we denote the intermediate Hilbert space prior to the projection (\ref{PaW}) by $\ch_{\rm phys}^{q|t}:=\ct_t(\ch_{\rm phys})$. It is clear that this is simply a different representation of the physical Hilbert space. 

In summary, evaluating the relational quantum Dirac observables relative to $t$ in the Dirac quantized theory produces entirely equivalent results to evaluating the reduced evolving observables in the Heisenberg picture of the reduced quantum theory in $t$ time.
It is, of course, well known that the Dirac quantized version of the parametrized particle is equivalent to its reduced quantization in $t$ time, e.g.\ see \cite{Rovelli:2004tv, Gambini:2000ht, bianca-lecnotes}. However, this explicit map from one to the other, using the method of quantum symmetry reduction through trivializing the constraints and subsequently projecting onto the classical gauge fixing conditions is new and fully elucidates the relation between the two quantum theories.

\subsection{From Dirac to reduced quantum theory in $q$ time}\label{sec_Dirac2q}

We now repeat this exercise, mapping the clock-neutral Dirac quantum theory via constraint trivialization and subsequent `projection' onto the classical $q=0$ gauge condition to the reduced quantum theories relative to the clock $q$, which reside in the left and right moving reduced Hilbert spaces $\ch_\pm$ of sec.\ \ref{sec_qqtime}. In particular, we will now map the {\it canonical} Dirac quantized theory to the {\it affine} reduced quantum theories on $\ch_\pm$. As can be expected from the previous discussion, this step is more involved. 
The procedure once more constitutes a quantum deparametrization through quantum symmetry reduction.

Define the {\it trivialization map}
\ba
\ct_q:=\ct_{q+}+\ct_{q-}\,,\q\q\q\q \ct_{q\pm}:=\exp\left(\pm i\,\hat{q}\,(\widehat{\sqrt{-p_t}}-\epsilon)\right)\,\theta(\mp\hat{p})\,,\label{trivialq}
\ea
which thanks to the theta function will separate the left and right moving modes (we use $\theta(0)=\f{1}{2}$). $\epsilon>0$ is here an arbitrary positive number, whose role will become clear momentarily. Notice that otherwise (\ref{trivialq}) is entirely analogous to (\ref{trivialt}), intuitively being the evolution generator in $q$ time, except that the clock still appears as an operator $\hat{q}$. (\ref{trivialq}) will again map to a novel representation $\ch_{\rm phys}^{t|q}:=\ct_q(\ch_{\rm phys})$ of the physical Hilbert space.

We also have to define the inverse of (\ref{trivialq}), i.e.\ $\ct_q^{-1}:\ch_{\rm phys}^{t|q}\rightarrow\ch_{\rm phys}$. It is 
\ba\label{trivialqinv}
\ct_q^{-1}:=\ct^{-1}_{q+}+\ct^{-1}_{q-}\,,\q\q\q\q \ct^{-1}_{q\pm}:=\exp\left(\mp i\,\hat{q}\,(\widehat{\sqrt{-p_t}}-\epsilon)\right)\,\theta(\mp\hat{p})\,.
\ea
Indeed, in appendix \ref{app_Dirac2q} we show that 
\ba
\ct_q^{-1}\,\ct_q = \theta(-\hat{p})+\theta(\hat{p}) = \mathds{1}\,,\q\text{on } \ch_{\rm phys}\,.\label{id}
\ea
We emphasize that this equation {\it only} holds on $\ch_{\rm phys}$, which is all we will need, and {\it only} for $\epsilon>0$. Hence, the parameter $\epsilon$ ensures that the trivialization map will be invertible.

The map (\ref{trivialq}) indeed trivializes the constraint $\hat{C}_H$ to the clock $q$, however, does so separately for the left and right moving sector. After a straightforward calculation one finds
\ba
\ct_q\,\hat{C}_H\,\ct_q^{-1}=\left(\hat{p}-2\,\widehat{\sqrt{-p_t}}+\epsilon\right)\left(\hat{p}+\epsilon\right)\,\theta(-\hat{p})+\left(\hat{p}+2\,\widehat{\sqrt{-p_t}}-\epsilon\right)\left(\hat{p}-\epsilon\right)\,\theta(\hat{p})\,,\label{trivialq2}
\ea
which is easy to interpret once recalling the factorization (\ref{factorize}), computing
\ba
\ct_q\,\hat{C}_\pm\,\ct_q^{-1} = (\hat{p}\pm\epsilon)\,\theta(\mp\hat{p})+\left(\hat{p}\pm2\,\widehat{\sqrt{-p_t}}\mp\epsilon\right)\,\theta(\pm\hat{p})\,,
\ea
and noting that\footnote{The second equality only holds once evaluated on $\ch_{\rm phys}^{t|q}$, so that the intermediate $\ct_q^{-1}\,\ct_q$ cancels thanks to (\ref{id}).}
\ba
\ct_q\,\hat{C}_H\,\ct_q^{-1}=\ct_q\,\hat{C}_+\,\hat{C}_-\,\ct_q^{-1}=\ct_q\,\hat{C}_+\,\ct_q^{-1}\,\ct_q\,\hat{C}_-\,\ct_q^{-1}\,.
\ea
That is, $\hat{C}_\pm$ gets trivialized to $\hat{p}\pm\epsilon$ on the left/right moving sector and this carries over to the according factorized trivialization of $\hat{C}_H$ in (\ref{trivialq2}).

Accordingly, applying $\ct_q$ to physical states represented as in (\ref{phys2}), the states on $\ch_{\rm phys}^{t|q}$ take the form
\ba
\ket{\psi}_{\rm phys}^{t|q}:=\ct_q\,\ket{\psi}_{\rm phys}=\int_{-\infty}^0\,\f{\mathrm{d}p_t}{2\sqrt{-p_t}}\Big(\psi_{\rm kin}(-\sqrt{-p_t},p_t)\,\ket{-\epsilon}_q\ket{p_t}_t+\psi_{\rm kin}(\sqrt{-p_t},p_t)\,\ket{+\epsilon}_q\ket{p_t}_t\Big)\,.
\ea
In this form, it is now also particularly evident why $\ct_q$ would fail to be invertible for $\epsilon=0$: one could no longer distinguish the left and right moving sectors. We thus keep $\epsilon>0$.  Again, the clock factor of the state has become essentially redundant, except for distinguishing the left and right moving sectors, which is why we shall not yet `project' it out. 

Using (\ref{PIP}), it is straightforward to check that
\ba
(\phi_{\rm phys},\psi_{\rm phys})_{\rm phys}={}_{\rm kin}\la\phi|\,\psi\ra_{\rm phys}={}_{\rm kin}\la\ct_q\phi|\,\psi\ra^{t|q}_{\rm phys}=(\phi_{\rm phys}^{t|q},\psi_{\rm phys}^{t|q})_{\rm phys}^{t|q}\,,
\ea
so that $\ct_q$ is an isometry from $\ch_{\rm phys}$ to $\ch_{\rm phys}^{t|q}$.

Next, we need to show that the regularized relational quantum Dirac observables (\ref{qTX}) transform correctly. Ultimately, we wish to reproduce the regularized evolving time-of-arrival observables (\ref{redqTX}) of the left and right moving reduced theories. This is the most non-trivial part of the procedure. As an intermediate step, we show in appendix \ref{app_toatrans} that the observables (\ref{qTX}) transformed to $\ch_{\rm phys}^{t|q}$ become
\ba
\ct_q\,\hat{T}_\delta(X)\,\ct_q^{-1}&=&\left(\hat{t}_\delta-\f{X}{2}\,\widehat{(\sqrt{-p_t})_\delta^{-1}}\right)\,\theta(-\hat{p})+\left(\hat{t}_\delta+\f{X}{2}\,\widehat{(\sqrt{-p_t})_\delta^{-1}}\right)\,\theta(\hat{p})+\f{i}{4}\widehat{(p_t)_\delta^{-1}}\,,\nn\\
\ct_q\,\hat{P}_T(X)\,\ct_q^{-1}&=&\hat{p}_t\,\theta(-\hat{p})+\hat{p}_t\,\theta(\hat{p})\,,\label{toatrans1}
\ea
where $\widehat{(p_t)_\delta^{-1}}$ is defined in (\ref{invpt2}) and, in analogy to (\ref{invroot}) on $\ch_\pm$, the regularized inverse square root is given by
\ba\label{invroot2}
\widehat{(\sqrt{-p_t})_\delta^{-1}}\ket{p_t}:=\begin{cases}
\f{1}{\sqrt{-p_t}}\ket{p_t}     & p_t\leq-\delta^2 , \\
   \f{\sqrt{-p_t}}{\delta^2}\ket{p_t}   & -\delta^2<p_t\leq0.
\end{cases}
\ea

That is, the observables on $\ch_{\rm phys}^{t|q}$ are already almost of the form as the regularized time-of-arrival observables (\ref{redqTX}) on $\ch_\pm$. The remaining differences can be easily traced back to the different representations and, in particular measures, which we use on $\ch_{\rm phys}^{t|q}$ and the affine $\ch_{\pm}$. Firstly, comparing (\ref{affpip}) and (\ref{PIP}), we see that
\ba
(\phi_{\rm phys},\psi_{\rm phys})_{\rm phys}=(\phi_{\rm phys}^{t|q},\psi_{\rm phys}^{t|q})_{\rm phys}^{t|q}=\f{1}{2}\,\la\phi\ket{\psi}_++\f{1}{2}\,\la\phi\ket{\psi}_-\,,\label{inp}
\ea
where $\la\phi\ket{\psi}_\pm$ is the inner product (\ref{affpip}) on $\ch_{\pm}$, provided we identify
\ba
\psi_\pm(p_t):=(-p_t)^{1/4}\,\psi_{\rm kin}(\mp\sqrt{-p_t},p_t)\,,\label{identify2}
\ea
where $\psi_\pm$ are the wave functions of the left and right moving modes on $\ch_\pm$, respectively. That is, in harmony with (\ref{PIP}), the physical inner product then equals half the sum of the inner products in the left and right moving Hilbert spaces $\ch_\pm$, consistent with the fact that all states can then be simultaneously normalized.

Recalling the normalization (\ref{unnorm}) of the reduced generalized momentum eigenstates ${}_\pm\bra{p_t}p_t'\ra_\pm=-p_t\,\delta(p_t-p_t')$ on $\ch_\pm$, while on $\ch_{\rm phys}^{t|q}$ we have inherited ${}_t\bra{p_t}p_t'\ra_t=\delta(p_t-p_t')$ from $\ch_{\rm kin}$, we can now write
\ba
\widehat{(-p_t)^{1/4}}\,\ct_q\,\ket{\psi}_{\rm phys}&=&-\int_{-\infty}^0\,\f{\mathrm{d}p_t}{2p_t}\Big(\psi_+(p_t)\,\ket{-\epsilon}_q\ket{p_t}_++\psi_-(p_t)\,\ket{+\epsilon}_q\ket{p_t}_-\Big)\,\nn\\
&=&\f{1}{2}\,\ket{-\epsilon}_q \ket{\psi}_++\f{1}{2}\,\ket{+\epsilon}_q\ket{\psi}_-\,,\label{lasttrans}
\ea
where we identify $\ket{\psi}_\pm$ with the reduced states (\ref{affstate}) on $\ch_\pm$. 

Consequently, we need to transform the relational observables (\ref{toatrans1}) further to have them act on the reduced states and check that they actually reproduce the reduced evolving observables (\ref{redqTX}). In appendix \ref{app_recover}, we prove that this is indeed the case, producing
\ba
\widehat{(-p_t)^{1/4}}\,\ct_q\,\hat{T}_\delta(X)\,\ct_q^{-1}\,\widehat{(-p_t)^{-1/4}}&=&\left(\hat{t}_{\delta+}-\f{X}{2}\,\widehat{(\sqrt{-p_t})_\delta^{-1}}\right)\,\theta(-\hat{p})+\left(\hat{t}_{\delta-}+\f{X}{2}\,\widehat{(\sqrt{-p_t})_\delta^{-1}}\right)\,\theta(\hat{p})\nn\\
&=&\hat{T}_+(X)\,\theta(-\hat{p})+\hat{T}_-(X)\,\theta(\hat{p})\,,\nn\\
\widehat{(-p_t)^{1/4}}\,\ct_q\,\hat{P}_T(X)\,\ct_q^{-1}\,\widehat{(-p_t)^{-1/4}}&=&\hat{p}_t\,\theta(-\hat{p})+\hat{p}_t\,\theta(\hat{p})\nn\\
&=&\hat{P}_{T_+}(X)\,\theta(-\hat{p})+\hat{P}_{T_-}(X)\,\theta(\hat{p})\,,\label{toatrans2}
\ea
where all operators in these expressions (except $\hat{p}$) now finally coincide with the corresponding reduced operators on $\ch_\pm$ of sec.\ \ref{sec_qqtime}. Notice that in these transformations $\widehat{(-p_t)^{-1/4}}$ is {\it not} regularized and defined by spectral decomposition. The reason it is not regularized is that it simply amounts to a measure factor in the integral representation of states and we also need it to render the transformation (\ref{lasttrans}) invertible. We have thereby finally recovered the reduced evolving observables in the corresponding left and right moving sectors from the relational Dirac observables (\ref{qTX}) on $\ch_{\rm phys}$. In particular, we have correctly mapped the regularized Dirac observables into the regularized reduced observables. This constitutes a non-trivial consistency check of the construction.

To complete the reduction to the affine reduced theories, we only have to `project' out the redundant clock factor in the state (\ref{lasttrans}) by projecting it onto the classical gauge fixing condition $q=0$, however, per sector. Indeed, in analogy to (\ref{PaW}), we get
\ba
\f{1}{\sqrt{2}}\,\ket{\psi}_{\pm}=2\sqrt{\pi}\,{}_q\bra{q=0}\,\theta(\mp\hat{p})\,\widehat{(-p_t)^{1/4}}\,\ct_q\,\ket{\psi}_{\rm phys}\,.\label{q0state}
\ea
It is clear that this `projection' is compatible with the observables (\ref{toatrans2}), `projecting' them to the correct ones on the Hilbert spaces $\ch_\pm$ of left and right moving modes. Likewise, this `projection' is compatible with the inner product (\ref{inp}); upon `projection', the reduced inner product (\ref{affpip}) provides equivalent results. 
The reduced left and right mover states $\ket{\psi}_\pm$ each comprise half of the information and normalization of the complete physical state. Both are needed to invert the transformation and recover a full physical state from the left and right moving sectors. Hence, we obtain two `quantum coordinate maps' in the form of quantum symmetry reduction maps relative to the left and right moving sectors of clock $\hat q$:
\ba
\Phi_\pm:=2\sqrt{\pi}\,{}_q\bra{q=0}\,\theta(\mp\hat{p})\,\widehat{(-p_t)^{1/4}}\,\ct_q\,.\label{Phipm}
\ea

Finally, we emphasize that the image of the complete transformation is, in analogy to sec.\ \ref{sec_Dirac2t}, the Heisenberg picture on $\ch_\pm$. In particular, the states (\ref{q0state}) can be regarded as initial states at $q=0$.

\subsection{Switching relational quantum clocks}\label{sec_qclockswitch}

We are now ready for the final step of the construction: switching between the relational quantum dynamics relative to $t$ and $q$. This is the quantum analog of the classical construction in sec.\ \ref{sec_clswitch}. It is clear how to proceed: we have just built the `quantum coordinate maps' $\Phi_t,\Phi_\pm$ from the clock-neutral physical Hilbert space $\ch_{\rm phys}$ to the reduced Hilbert spaces $\ch_{q|t}$ and $\ch_\pm$ relative to the clocks $t$ and $q$, respectively. We just have to appropriately invert a given reduction map and concatenate it with the other. Notice that this means in particular that we will always change quantum clocks {\it via} the clock-neutral physical Hilbert space, just like we changed classical clocks in sec.\ \ref{sec_clswitch} via the clock-neutral constraint surface. Clock changes thus assume the compositional form of `quantum coordinate changes'. We emphasize that the quantum coordinate maps can be inverted, despite the `projection' onto classical gauge-fixing conditions contained in them. The reason is that, as stated before, it is not a true projection when applied to the physical Hilbert space, as it only removes redundant information (it is a projection when applied to $\ch_{\rm kin}$).

We begin by switching from the quantum evolution relative to $t$ to that relative to $q$ and wish to construct a map
$
\hat{\cs}_{t\to q\pm}:\ch_{q|t}\rightarrow\ch_\pm\,.
$
Inverting the reduction map of sec.\ \ref{sec_Dirac2t} and concatenating it with the reduction map of sec.\ \ref{sec_Dirac2q} yields:
\ba
\hat{\cs}_{t\to q\pm}&:=&\Phi_\pm\circ\Phi_t^{-1}\nn\\
&=&2\sqrt{\pi}\,{}_q\bra{q=0}\,\theta(\mp\hat{p})\,(-\hat{p}_t)^{1/4}\,\ct_q\,\ct_t^\dag\,\ket{p_t=0}_t\otimes [\cdot]_{q|t}\,.\label{t2q}
\ea
By the term $\ket{p_t=0}_t\otimes [\cdot]_{q|t}$ we mean inserting the reduced state $\ket{\psi}_{q|t}\in\ch_{q|t}$ into the empty slot and tensoring it with the zero-clock-momentum state.\footnote{For notational simplicity, we had suppressed tensor products in our description so far. It should be emphasized that it is a kinematical tensor product, which ultimately comes from decomposing the kinematical Hilbert space as $\ch_{\rm kin}\simeq\ch_t\otimes\ch_q$, where $\ch_t$ corresponds to the kinematical Hilbert space of the pair $(\hat t,\hat p_t)$ and $\ch_q$ to the kinematical Hilbert space of the pair $(\hat q,\hat p)$.} \footnote{The reader may be concerned that there is an ambiguity in how to embed reduced quantum states into the physical Hilbert space. Mathematically, this is of course true. However, this ambiguity is resolved (up to unitary equivalence on the reference system Hilbert space) when keeping in mind the physical interpretation of the reduced state as the description of the evolving degrees of freedom relative to the reference system of choice. In particular, the reduced state should be interpreted as the state obtained through applying the corresponding `quantum coordinate map' to a physical state and this information should not be discarded if one wants to invert the map. This is qualitatively the same as in general relativity when changing coordinates. Also there a coordinate description relative to a choice of frame comes with an interpretation as the image of the spacetime physics through a coordinate map. Without this one could not correctly associate coordinates to points in the manifold and in turn not change coordinate frame perspectives. 

This relates to our earlier comment that in general we give primacy to Dirac quantization and consider the symmetry reduced theory as the derived structure with an accordingly induced interpretation. It is only a coincidence that in the present model, the quantum symmetry reduced theory coincides with the quantization of the classically symmetry reduced theory and could thereby also be given a standalone interpretation. } Since $\ket{p_t=0}_t=1/\sqrt{2\pi}\,\int\,\mathrm{d}t\,\ket{t}_t$, this step, in fact, corresponds precisely to averaging over the classical gauge fixing conditions $t=const$ and thereby to restoring the gauge invariance of the system. 
It is easy to see from the discussion in secs.\ \ref{sec_qttime}--\ref{sec_Dirac2q} that we immediately have
\ba
\hat{\cs}_{t\to q\pm}\,\ket{\psi}_{q|t}=\hat{\cs}_{t\to q\pm}\,\int_{-\infty}^\infty\,\mathrm{d}p\,\psi_{q|t}(p)\,\ket{p}_q=\f{1}{\sqrt{2}}\,\ket{\psi}_\pm=\f{1}{\sqrt{2}}\,\int_{-\infty}^0\,\f{\mathrm{d}p_t}{-p_t}\,\psi_\pm(p_t)\,\ket{p_t}_\pm\,,
\ea
if one invokes the identifications (\ref{identify}, \ref{identify2}),
\ba
\psi_{q|t}(p)\equiv\psi_{\rm kin}(p,-p^2)\,,\q\q\q\q\psi_\pm(p_t)\equiv(-p_t)^{1/4}\,\psi_{\rm kin}(\mp\sqrt{-p_t},p_t)\,.\label{identify3a}
\ea
Indeed, in this manner, both reduced states correspond to the {\it same} physical state $\ket{\psi}_{\rm phys}$, defined through (\ref{phys1}, \ref{phys2}). Specifically, while the identification is done in terms of a kinematical wave function, it actually applies to the entire equivalence class of those kinematical states that map to the same physical state, so that no ambiguity arises. This provides a unique map from $\ch_{q|t}$ {\it via} $\ch_{\rm phys}$ to $\ch_\pm$.

As shown in appendix \ref{app_t2q}, the map (\ref{t2q}) is equivalent to a direct map between $\ch_{q|t}$ and $\ch_\pm$
\ba
\hat{\cs}_{t\to q\pm} \equiv \sqrt{2}\,\hat{\cp}_{q\to t}\,\theta(\mp\hat{p})\,\widehat{\sqrt{|p|}}\,,
\ea
and can thereby be simplified and expressed solely in terms of operators on the reduced Hilbert spaces.
Here, in some analogy to the parity-swap of \cite{Giacomini:2017zju} (see also \cite{Vanrietvelde:2018pgb}), we have defined the swap operator
\ba
\hat{\cp}_{q\to t}\,\ket{p}_q:=\f{1}{|p|}\,\ket{-p^2}_\pm\,.\label{swap1}
\ea
Notice that this map from the (generalized) $q$-momentum eigenstates on $\ch_{q|t}$ to the (generalized) $t$-momentum eigenstates on $\ch_\pm$ respects their different normalizations. 

It is convenient to summarize these maps in the following commutative diagram (cf.\ the classical clock change maps in sec.~\ref{sec_clswitch}): 
\begin{center}
\begin{tikzcd}
& \mathcal{H}_{\textrm{phys}}  \arrow[rd, "\mathcal{T}_q"]& \\
\mathcal{H}^{q|t}_{\textrm{phys}} \arrow[ru, "\mathcal{T}_t^\dagger"] & & \mathcal{H}^{t|q}_{\textrm{phys}} \arrow[d, "2 \sqrt{\pi}\,_q\bra{q=0} \theta(\mp \hat{p}) (- \hat{p}_t)^\frac{1}{4}"]\\
\mathcal{H}_{q|t} \arrow[rr, "\hat{\mathcal{S}}_{t \to q \pm}"] \arrow[u, "\ket{p_t = 0}_t\otimes (\cdot)_{q|t}   "]  & & \mathcal{H}_{\pm}
\end{tikzcd}
\end{center}
This makes it explicit that the construction of the quantum clock switch proceeds via the clock-neutral physical Hilbert space, underscoring the discussion in \cite{Vanrietvelde:2018pgb, Vanrietvelde:2018dit}. In particular, the clock change (\ref{t2q}) takes the compositional form $\Phi_\pm\circ\Phi_t^{-1}$ of a `quantum coordinate change'  where the clock-neutral $\ch_{\rm phys}$ plays the role of the `quantum manifold'.

We can analogously construct the inverse switch $\hat{\cs}_{q\pm\to t}:\ch_\pm\rightarrow \ch_{q|t}$ from the quantum relational dynamics relative to $q$ to that relative to $t$ by
\ba
\hat{\cs}_{q\pm\to t}&:=&\Phi_t\circ\Phi_\pm^{-1}\nn\\
&=&\sqrt{2\pi}\,{}_t\bra{t=0}\,\ct_t\,\ct_q^{-1}\,\widehat{(-p_t)^{-1/4}}\,\f{\ket{p=\mp\epsilon}_q}{2}\,\otimes [\cdot]_\pm\nn\\
&=&\sqrt{2\pi}\,{}_t\bra{t=0}\,\ct_t\,\ct_{q\pm}^{-1}\,\widehat{(-p_t)^{-1/4}}\,\f{\ket{p=\mp\epsilon}_q}{2}\,\otimes [\cdot]_\pm\,,\label{q2t}
\ea
where $[\cdot]_\pm$ means that we have to insert the reduced state $\ket{\psi}_\pm\in\ch_\pm$ into it and tensor it with the rest.
From the discussion in secs.\ \ref{sec_qttime}--\ref{sec_Dirac2q} it immediately follows that
\ba
\hat{\cs}_{q\pm\to t}\,\ket{\psi}_\pm&=&\sqrt{2\pi}\,{}_t\bra{t=0}\,\ct_t\,\theta(\mp\hat{p})\,\ket{\psi}_{\rm phys}\nn\\
&=&\theta(\mp\hat{p})\,\int_{-\infty}^\infty\,\mathrm{d}p\,\psi_{q|t}(p)\,\ket{p}_q=\theta(\mp\hat{p})\,\ket{\psi}_{q|t}\,,
\ea
again, once invoking the identifications (\ref{identify3a}). This provides unique maps from $\ch_\pm$ to $\ch^\pm_{q|t}:=\theta(\mp\hat{p})(\ch_{q|t})$ {\it via} $\ch_{\rm phys}$, i.e.\ from the left/right mover spaces $\ch_\pm$ to the left/right moving sector of $\ch_{q|t}$, respectively. Hence, each map yields only half of the reduced quantum theory in $t$ time.

In appendix \ref{app_t2q} it is demonstrated that the map (\ref{q2t}) can also be simplified and expressed entirely in terms of properties of the reduced Hilbert spaces, being equivalent to
\ba
\hat{\cs}_{q\pm\to t}\equiv \f{1}{2}\,\hat{\cp}_{t\to q\pm}\,\widehat{(-p_t)^{-1/4}}\,,
\ea
where, in analogy to (\ref{swap1}), we have defined the swap
\ba
\hat{\cp}_{t\to q\pm}\,\ket{p_t}_\pm:=\sqrt{-p_t}\,\ket{\mp\sqrt{-p_t}\,}_q\,.
\ea
Again, this map from the affine (generalized) $t$-momentum eigenstates on $\ch_\pm$ to the (generalized) $q$-momentum eigenstates on $\ch_{q|t}$ respects their different normalizations. Specifically, note that
\ba
\hat{\cp}_{t\to q\pm}\cdot\hat{\cp}_{q\to t}=\mathds{1}_{\ch^\pm_{q|t}}\,,\q\q\q\q \hat{\cp}_{q\to t}\cdot\hat{\cp}_{t\to q\pm}=\mathds{1}_{\ch_\pm}\,.
\ea
In summary, we have the commutative diagram (cf.\ the corresponding classical clock change map in sec.~\ref{sec_clswitch}):
\begin{center}
\begin{tikzcd}
& \mathcal{H}_{\textrm{phys}}  \arrow[rd, "\mathcal{T}_t"]& \\
\mathcal{H}^{t|q}_{\textrm{phys}} \arrow[ru, "\mathcal{T}_{q \pm}^{-1}"] & & \mathcal{H}^{q|t}_{\textrm{phys}} \arrow[d, "2 \sqrt{\pi}\,_t\bra{t=0}"]\\
\mathcal{H}_{\pm} \arrow[rr, "\hat{\mathcal{S}}_{q \pm \to t}"] \arrow[u, "\widehat{(- p_t)^{-\frac{1}{4}}} \frac{1}{2}\ket{p = \mp \epsilon}_q \otimes(\cdot)_\pm"]  & & \mathcal{H}^{\pm}_{q|t} \\
\end{tikzcd}
\end{center}
This constitutes the `quantum coordinate transformation' $\Phi_t\circ\Phi_\pm^{-1}$ via the clock-neutral `quantum manifold' $\ch_{\rm phys}$.

Having now constructed the quantum clock switches in both directions, we are also in the position to check how observables transform between the reduced theories. We begin by mapping the elementary observables $\hat{q},\hat{p}$ from $\ch_{q|t}$ to $\ch_\pm$. After some straightforward calculations one finds
\ba
\hat{\cs}_{t\to q\pm}\,\hat{q}\,\hat{\cs}_{q\pm\to t} = \mp\left(\widehat{(\sqrt{-p_t})^{-1}}\,\hat{\ft}+\hat{\ft}\,\widehat{(\sqrt{-p_t})^{-1}}\right)\,\f{1}{\sqrt{2}}\,,\q\q\q\q\hat{\cs}_{t\to q\pm}\,\hat{p}\,\hat{\cs}_{q\pm\to t} =\mp\widehat{\sqrt{-p_t}}\,\f{1}{\sqrt{2}}\,. 
\ea
Recall that classically $t=\ft/p_t$ and notice the similarity with the classical transformation (\ref{clq2t}). The factor $1/\sqrt{2}$ here comes from the normalization conditions (\ref{inp}, \ref{q0state}). The reader might wonder why we do not have {\it regularized} inverse square root operators appearing in the left equation. The reason is that the image of $\hat{\cs}_{q\pm\to t}$ on the left hand side of the equation is $\ch_{q|t}^\pm$, i.e.\ only half of $\ch_{q|t}$ and $\hat{q}$ does {\it not} act as a self-adjoint operator on this subset; being conjugate to $\hat{p}$, it can map states in $\ch_{q|t}^+$ to states in $\ch_{q|t}^-$ and vice versa. Hence, one would actually have to regularize $\hat{q}$ on $\ch_{q|t}^\pm$ to produce a self-adjoint operator on the right hand side also. It is clear that this can be done, however, we refrain from doing so.

In particular, we can now map the reduced evolving observables (\ref{redqQT}) from $\ch_{q|t}$ to $\ch_\pm$:
\ba
\hat{\cs}_{t\to q\pm}\,\hat{Q}(\tau)\,\hat{\cs}_{q\pm\to t} &=& \mp\left(\widehat{(\sqrt{-p_t})^{-1}}\,\hat{\ft}+\hat{\ft}\,\widehat{(\sqrt{-p_t})^{-1}}+2\,\widehat{\sqrt{-p_t}}\,\tau \right)\,\f{1}{\sqrt{2}}=:\hat{Q}_\pm(\tau)\,\f{1}{\sqrt{2}}\,,\nn\\
\hat{\cs}_{t\to q\pm}\,\hat{P}(\tau)\,\hat{\cs}_{q\pm\to t} &=&\mp\widehat{\sqrt{-p_t}}\,\f{1}{\sqrt{2}}\,. \label{Qpm}
\ea
Notice the similarity with the classical expression in (\ref{onboth}).

Conversely, we can map the elementary operators from $\ch_\pm$ to $\ch_{q|t}^\pm$. After some computation one finds
\ba
\hat{\cs}_{q\pm\to t}\,\hat{\ft}\,\hat{\cs}_{t\to q\pm} = \f{1}{4}\,\left(\hat{q}\,\hat{p}+\hat{p}\,\hat{q}\right)\,\theta(\mp\hat{p})\,\f{1}{\sqrt{2}}\,,\q\q\q\q \hat{\cs}_{q\pm\to t}\,\hat{p}_t\,\hat{\cs}_{t\to q\pm} = -\hat{p}^2\,\theta(\mp\hat{p})\,\f{1}{\sqrt{2}}\,.
\ea
This is the quantum analog of the classical relation in (\ref{clt2q}). Specifically, note that in the left equation we are transforming $\ft$, which classically equals $t\,p_t$. Furthermore, it is straightforward to check that the reduced evolving observables (\ref{redqTX}) map from $\ch_\pm$ to $\ch_{q|t}^\pm$ as follows
\ba
\hat{\cs}_{q\pm\to t}\,\hat{T}_\pm(X)\,\hat{\cs}_{t\to q\pm} &=& \f{1}{2}\,\left(X\,\widehat{(p)^{-1}_\delta}-\f{1}{2}\left(\widehat{(p)^{-1}_\delta}\,\hat{q}+\hat{q}\, \widehat{(p)^{-1}_\delta}\right)\right)\,\theta(\mp\hat{p})\,\f{1}{\sqrt{2}}\,,\nn\\
\hat{\cs}_{q\pm\to t}\,\hat{P}_{T_\pm}(X)\,\hat{\cs}_{t\to q\pm} &=&-\hat{p}^2\,\theta(\mp\hat{p})\,\f{1}{\sqrt{2}}\,.
\ea
Here, we recover the correctly regularized operators also on $\ch_{q|t}$. Notice again the similarity to the classical expression in (\ref{onboth}).

\subsection{Application: Quantum relativity of comparing clock readings}\label{sec_qrelativity}

Lastly, we consider how to switch from final to initial clock `readings' when changing the quantum clock, so we can consistently continue the relational dynamics afterwards. We recall from sec.\ \ref{sec_clswitch} that when classically changing from $t$ to $q$ time, we used $X_i=Q_\pm(\tau_f)$ for the initial `reading' of clock $q$ after the switch. Similarly, after switching from $q$ to $t$ time, we used $\tau_i=T(X_f)$ as the initial `reading' of $t$ for the continued evolution. It is clear that we cannot na\"ively use the same procedure in the quantum theory as $\hat{Q}_\pm(\tau_f),\hat{T}(X)$ are now operators. Instead, the natural quantum analog is to exploit the quantum state, which we now know how to transform and to set initial clock readings after the switch in terms of expectation values:
\ba
X_i:=\la\hat{Q}_\pm(\tau_f)\ra_\pm\,,\q\q\q\q\q \tau_i:=\la\hat{T}(X_f)\ra_{q|t}\,.\label{jump0}
\ea

In contrast to the classical case, this will in general not produce a continuous evolution from one clock to the other. In particular, in the quantum theory we will generally have
\ba
\Big\la\,\hat{T}_\pm(\la\hat{Q}_\pm(\tau_f)\ra_\pm)\,\Big\ra_\pm\neq\tau_f\,,\q\q\q\q\Big\la\,\hat{Q}(\la\hat{T}(X_f)\ra_{q|t})\,\Big\ra_{q|t}\neq X_f\,,\label{jump1}
\ea
because of quantum uncertainties. To see this, we employ (\ref{redqTX}), the left definition in (\ref{jump0}), (\ref{Qpm}) and (\ref{regtime0}) to find
\ba
\Big\la\,\hat{T}_\pm(X_i)\,\Big\ra_\pm &=& \Big\la\,\hat{T}_\pm(\la\hat{Q}_\pm(\tau_f)\ra_\pm)\,\Big\ra_\pm\nn\\
&=&\f{\la\widehat{(p_t)_\delta^{-1}}\,\hat{\ft}+\hat{\ft}\,\widehat{(p_t)_\delta^{-1}}\ra_\pm}{2}+\f{\la\widehat{(\sqrt{-p_t})^{-1}}\,\hat{\ft}+\hat{\ft}\,\widehat{(\sqrt{-p_t})^{-1}}\ra_\pm}{2}\,\la \widehat{(\sqrt{-p_t})_\delta^{-1}}\ra_\pm\nn\\
&&\q\q\q\q +\la\widehat{\sqrt{-p_t}}\ra_\pm\,\la \widehat{(\sqrt{-p_t})_\delta^{-1}}\ra_\pm\,\tau_f\,.\nn
\ea
It is clear that for general reduced states $\ket{\psi}_\pm$ we have, for example, $\la\widehat{\sqrt{-p_t}}\ra_\pm\,\la \widehat{(\sqrt{-p_t})_\delta^{-1}}\ra_\pm\neq1$ so that the r.h.s.\ of the equation cannot be equal to $\tau_f$. This yields the first discrepancy in (\ref{jump1}) and is independent of the fact that inverse powers of the $t$-momentum appear in both regularized and non-regularized form. Similarly, one shows the second discrepancy in (\ref{jump1}).

To understand this more intuitively, note that the equal time surfaces for the two clocks correspond to orthogonal hypersurfaces in the extended configuration space $\mathbb{R}^2$ and the physical state in $\ch_{\rm phys}$, corresponding to the reduced states in question, will generally have a different spread along the two orthogonal directions. Heuristically, a reduced quantum state $\ket{\psi}_{q|t}$ can be thought of as a section of the physical state on a $t=const$ surface in configuration space, while a reduced state $\ket{\psi}_\pm$ corresponds to a section of the same physical state on a $q=const$ surface. The physical state having different spreads in different directions implies that the reduced wave functions $\psi_{q|t}$ and $\psi_\pm$ will generally feature different spreads. As such, the clock expectation values computed in the two reduced theories will generally lead to the disagreement (\ref{jump1}). 

Said another way, if clock $\hat q$ reads $X_i$ relative to clock $\hat t$ when the latter reads $\tau_f$, then generally clock $\hat t$ will \emph{not} read $\tau_f$ relative to clock $\hat q$ when the latter reads $X_i$. Considering the two clocks as defining two different temporal reference frames, we thus obtain a relativity of clock comparisons: there is no absolute, i.e.\ temporal-frame-independent way to say what the `simultaneous' readings of different clocks are. This is a consequence of the fact that the evolutions of $\hat q$ relative to $\hat t$ and, conversely, of $\hat t$ relative to $\hat q$ correspond to two \emph{distinct} families of relational Dirac observables $\hat Q(\tau)$ and $\hat T_\pm(X)$ on the clock-neutral physical Hilbert space $\ch_{\rm phys}$.

In line with the above, clock changes have been thoroughly analyzed in a quantum phase space language in a semiclassical regime, exhibiting jumps of order $\hbar$ in expectation values when switching from the evolution of one clock to another \cite{Bojowald:2010xp,Bojowald:2010qw, Hohn:2011us}. In a forthcoming article \cite{pqps}, it will be shown that the full quantum method of this article is equivalent to the effective clock changes of \cite{Bojowald:2010xp,Bojowald:2010qw, Hohn:2011us}, once restricted to the semiclassical regime.

\section{Conclusions and outlook}\label{sec_conc}

By means of the parametrized particle, we have displayed a systematic method for switching between different choices of relational quantum clocks. This extends the recent approach to switching quantum reference systems \cite{Vanrietvelde:2018pgb, Vanrietvelde:2018dit,Giacomini:2017zju,FEC}, which thus far was only applied to spatial quantum reference frames, to the temporal case, underscoring its unifying character. Since the new method is fully quantum and developed directly at a Hilbert space level, our work also constitutes an extension of the semiclassical method of clock changes \cite{Bojowald:2010xp,Bojowald:2010qw, Hohn:2011us}, which, moreover, was formulated in the language of quantum phase spaces. In forthcoming work \cite{pqps}, the equivalence of the two methods will be established, once restricting the novel method exhibited here and in \cite{Vanrietvelde:2018pgb, Vanrietvelde:2018dit} to the semiclassical regime.

While here we have chosen a particularly simple toy model, it nevertheless showcases a surprisingly non-trivial behavior when choosing the particle's position as a quantum clock. In particular, we needed to carefully regularize the time-of-arrival observable in both the reduced and Dirac quantization in order to obtain a self-adjoint observable that admits the interpretation of a genuine quantum observable, incl.\ a probabilistic interpretation. Remarkably, our quantum symmetry reduction method correctly maps the regularized time-of-arrival observable from the canonical Dirac theory to its regularized form on the affine reduced theory relative to clock $q$, where in both cases we have used a symmetric operator ordering. This constitutes a non-trivial consistency check of our construction. We note that there is a debate in the literature about the physical interpretation of regularized time-of-arrival operators and, in fact, arguments that these cannot correspond to continuously monitoring the point of arrival in a laboratory \cite{aharonov1998measurement}, but see also \cite{Grot:1996xu,muga2000arrival}. This is, however, not a matter of concern for us here, our ambition being to demonstrate how non-trivial self-adjoint relational observables consistently transform between the different quantum theories, rather than considering the parametrized particle as a real physical system.

The quantum reduction method developed here and in \cite{Vanrietvelde:2018pgb, Vanrietvelde:2018dit}, and which is key to the systematic switches of quantum reference systems, is general: namely, (i) choose a quantum reference system in the perspective-neutral Dirac quantum theory, (ii) trivialize the constraint(s) to the degrees of freedom of the reference system, which then become redundant, and finally (iii) `project' onto the classical gauge fixing conditions, corresponding to the choice of reference system, to remove the redundant degrees of freedom. In \cite{pacosmo}, this is also confirmed in a simple quantum cosmological model where, again, one can consistently switch between different internal time functions and a novel perspective on the wave function of the universe ensues from the constructed quantum covariance. Furthermore, in the later works \cite{Hoehn:2019owq,Hoehn:2020epv} this method is generalized to various classes of models with a Hamiltonian constraint and an equivalence with the Page-Wootters formalism is established. 

We have also touched on the relation between Dirac and reduced quantization. In the simple model studied here the quantum symmetry reduced theory coincides with the quantization of the classically symmetry reduced model. However, as we have argued here and in \cite{Vanrietvelde:2018pgb, Vanrietvelde:2018dit,Hoehn:2019owq,Hoehn:2020epv}, on account of the generic inequivalence between Dirac and reduced quantization \cite{Ashtekar:1982wv,Ashtekar:1991hf,Schleich:1990gd,Kunstatter:1991ds,Hajicek:1990eu,Romano:1989zb,Dittrich:2016hvj,Dittrich:2015vfa,Loll:1990rx,Kaminski:2009qb,Domagala:2010bm,Giesel:2016gxq}, this will not be the case in more complicated models. Our novel quantum symmetry reduction method thereby suggests to rephrase the slogan ``constraining and quantizing don't commute" somewhat more precisely as  ``symmetry reduction and quantization don't commute".

In this article, we have profited from the clock $t$ being globally monotonic and the clock $q$ being well-behaved everywhere except at $p=0$. Of course, in more general systems, especially in the presence of interactions between the clocks \cite{Hohn:2011us,Marolf:1994nz,Marolf:2009wp, Giddings:2005id}, this will become substantially more complicated and clocks will feature turning points \cite{Kuchar:1991qf,Isham:1992ms,Anderson:2017jij,Bojowald:2010xp,Bojowald:2010qw, Hohn:2011us,Dittrich:2016hvj,Dittrich:2015vfa,Hajicek:1986ky, Hajicek:1994py,Hajicek:1995en,Hajicek:1988he,Schon:1989pe,Hajicek:1989ex,Rovelli:1990jm}. Indeed, as already emphasized, reduced quantum theories will fail to be globally valid in generic systems, in analogy to the Gribov problem in gauge theories, and so globally valid descriptions of the  physics {\it relative} to a quantum reference system will generally fail to exist. Under such circumstances, it is clear that our method cannot globally relate the descriptions relative to different quantum reference systems. However, as already shown in \cite{Vanrietvelde:2018dit} for the relational $N$-body problem, our method can consistently cope with such situations, providing non-global changes of perspective (for transient changes of relational clocks, see also \cite{Bojowald:2010xp,Bojowald:2010qw, Hohn:2011us}).

We thus propose this as a general perspective (on perspectives) in quantum cosmology and quantum gravity: to define a {\it complete} relational quantum theory as the conjunction of the quantum-reference-system-neutral Dirac quantized theory and the multitude of quantum symmetry reduced theories, corresponding to the different choices of quantum reference system. In particular, we propose this as the path to establishing a genuine quantum notion of general covariance in quantum gravity, which means to be able to consistently switch between arbitrary choices of quantum reference systems, each of which can be used as a vantage point to describe the physics of the remaining degrees of freedom. If successful, the multiple choice problem would thereby turn into a multiple choice {\it feature} of the complete relational quantum theory, just like the possibility to choose arbitrary reference frames in general relativity is one of its celebrated features. In particular, this also links with the diffeomorphism invariance in quantum gravity: the diffeomorphism-invariant physical Hilbert space of the Dirac quantized theory, e.g.\ in loop quantum gravity \cite{Rovelli:2004tv,Thiemann:2007zz, bianca-lecnotes}, would then assume the role of the perspective-neutral quantum structure, via which {\it all} the reduced quantum theories, corresponding to all possible quantum reference systems could be consistently linked, in some analogy to coordinate changes on a manifold \cite{Vanrietvelde:2018pgb} (see also \cite{Giacomini:2017zju,FEC} for a more operational perspective on quantum covariance). Just like coordinate changes, these changes of perspective need not be globally valid, as discussed above.

\section*{Acknowledgements}
PH thanks Bianca Dittrich for discussion about the time-of-arrival concept. The project leading to this publication has initially received funding from the European Union's Horizon 2020 research
and innovation programme under the Marie Sklodowska-Curie grant agreement No 657661 (awarded to PH). PH also acknowledges support through a Vienna Center for Quantum Science and Technology Fellowship. This work was supported in part by funding from Okinawa Institute of Science and Technology Graduate University.

\appendix

\section{Action and constraint for the parametrized free particle}\label{app_nrpart}

The action of the (unparametrized) non-relativistic free particle with configuration space $\cq=\mathbb{R}$ reads
\ba
S=\int\,\mathrm{d}t\,\f{m}{2}\dot{q}^2\,,\label{sunpar}
\ea
where a $ \dot{} $ denotes a derivative with respect to (absolute) time $t$, which here is external and thus non-dynamical. Note that the Lagrangian is a function on $T\cq=\mathbb{R}^2$, the space of positions and velocities. Henceforth, we shall set $m=1/2$ for later convenience, so we do not have to carry around factors of $m/2$ in the canonical formulation.

Our aim is to promote $t$ to a {\it dynamical} variable on an extended configuration space $\cq_{\rm ext}:=\mathbb{R}^2$, coordinatized by $(t,q)$, in such a way that the solutions $(t(s),q(s))$, in a new evolution parameter $s$ and following from an extended action principle, are equivalent to the solutions $q(t)$ following from the original action (\ref{sunpar}). It is already clear that the extended system must be subject to a reparametrization symmetry, since changing the new evolution parameter $s\mapsto \tilde{s}(s)$ will change the parametrization of the orbit $(t(\tilde{s}(s)),q(\tilde{s}(s)))$ in $\cq_{\rm ext}$, but not the orbit and, in particular, not the relations between $t$ and $q$ along that orbit. Indeed, the correct action for the extended system is given by \cite{Rovelli:2004tv, Tambornino:2011vg, bianca-lecnotes, Henneaux:1992ig}
\ba
S_{\rm ext}=\int\,\mathrm{d}s\,L_{\rm ext}(q,t,q',t')=\int\,\mathrm{d}s\,L\left(q,\f{q'}{t'}\right)t'=\int\,\mathrm{d}s\,\f{1}{4}\left(\f{q'}{t'}\right)^2t'\,,
\ea
where $L(q,\dot{q})=1/4\,\dot{q}^2$ is the original, unextended Lagrangian and a $'$ denotes differentiation with respect to $s$. Notice that $L_{\rm ext}$ is now a function on $T\cq_{\rm ext}=\mathbb{R}^4$. Invariance of $S_{\rm ext}$ under reparametrizations $s\mapsto \tilde{s}(s)$ is manifest. Given the form of the action, it is also clear that $S_{\rm ext}$ will take the same values on a path $(t(s),q(s))$ with fixed initial $(t_i=t(s_i),q_i=q(s_i))$ and final $(t_f=t(s_f),q_f=q(s_f))$ as the original action (\ref{sunpar}) does on a path $q(t)$ with fixed initial $q_i=q(t_i)$ and final $q_f=q(t_f)$.

Upon Legendre transforming to the extended phase space $T^*\cq_{\rm ext}=\mathbb{R}^4$, we find
\ba
p&=&\f{\p L_{\rm ext}}{\p q'}=\f{1}{2}\f{q'}{t'}\,\nn\\
p_t&=&\f{\p L_{\rm ext}}{\p t'}= -\f{1}{4}\left(\f{q'}{t'}\right)^2=-p^2\,.\label{legmom}
\ea
Note that the last equation yields the primary constraint 
\ba
C_H=p_t+p^2\approx0\,,\label{hamconnr}
\ea 
i.e.\ the Legendre transformation is not surjective and maps to a constraint surface $\cc$ in $T^*\cq_{\rm ext}$, defined by (\ref{hamconnr}), where the momenta are not independent. Here $\approx$ denotes a weak equality, i.e.\ an equality which only holds on $\cc$.

In fact, the extended Hamiltonian is up to a factor equal to this constraint because
\ba
H_{\rm ext}:=p_t\,t'+p\,q'-L_{\rm ext}=t'\,C_H,
\ea
which is why we have added the index $H$ to emphasize that it is a Hamiltonian constraint, which also generates the dynamics. There is thus no secondary constraint.

In standard systems, a Hamiltonian does not depend on velocities. However, the appearance of $t'$ here reflects the reparametrization invariance of the system and the fact that we now have a whole plethora of non-dynamical evolution parameters $s$ to choose from. In fact, we can consider $t'$ as an arbitrary factor. Thanks to (\ref{legmom}), the relation between the velocities $(t',q')$ and the momenta $(p_t,p)$ is many-to-one, as we can rescale the velocities by an arbitrary non-vanishing factor without changing the image of the Legendre transformation. It is therefore consistent to simply replace $t'$ by an arbitrary non-vanishing factor $N$, usually called the {\it lapse function}, so that we really have $H_{\rm ext}=N\,C_H$ and the arbitrariness in the lapse accounts for the arbitrariness in the parametrization of the dynamics. 
Indeed, the dynamics of some function $F$ on the constraint surface $\cc$ reads
\ba
\f{\mathrm{d} F}{\mathrm{d}s}= \{F,H_{\rm ext}\}\approx N\,\{F,C_H\}\,,
\ea
and different choices of lapse $N$ amount to considering the dynamics in different parametrizations $s$. 

 In particular, choosing $N=1$, as we shall do in the main text, amounts to choosing a parametrization such that $t$ grows linearly in $s$. Hence, $C_H$ is the generator of changes in a parametrization $s$ in which $t$ grows linearly. Yet, since $C_H$ thereby generates dynamics in a non-dynamical parameter $s$, we cannot directly interpret this as the physical motion because all physical information must reflect the gauge invariance of the action $S_{\rm ext}$ and should thus be independent of the parametrization. The physical, i.e.\ reparametrization-invariant motion will thus be encoded in the {\it relations} among the variables $t,q,p_t,p$, and we shall discuss this in detail in the main text. There we will also see that the relational dynamics will be fully equivalent to the original unextended dynamics of (\ref{sunpar}), once choosing $t$ as a 'clock'.


\section{Classical gauge transformation for clock switches}\label{app_clswitch}

In coordinates, the embedding map of the reduced phase space in $t$-time into the constraint surface reads 
\ba
\iota_{q|t}:\cp_{q|t}\hookrightarrow\cc\,,\q\q\q\q
(q,p)\mapsto \left(q,p,t=0,p_t=-p^2\right)
\ea
and has the image $\cc\cap\cg_{t=0}$. Conversely, the projection
\ba
\pi_{t=0}:\cc\cap\cg_{t=0}\rightarrow\cp_{q|t}\,,\q\q\q\q \left(q,p,t=0,p_t=-p^2\right)\mapsto (q,p)\,,
\ea
drops all redundant information and we have $\pi_{t=0}\circ\iota_{q|t}=\text{Id}_{\cp_{q|t}}$. Analogously, one constructs 
\ba
\iota_\pm:\cp_{\pm}&\hookrightarrow&\cc_\pm\subset\cc\,,\q\q\q\,\,\,\q\q\q\q\q\q\q (t,p_t)\mapsto(t,p_t,q=0,p=\mp \sqrt{-p_t})\,,\nn\\
\pi_\pm:\cc_\pm&\rightarrow&\cp_\pm\,,\q\q\q\q (t,p_t,q=0,p=\mp \sqrt{-p_t})\mapsto(t,p_t)\,.
\ea

It is easy to derive the gauge transformation that takes $\cc\cap\cg_{t=0}$ to $\cc\cap\cg_{q=0}$. Let $\alpha_{C_H}^s$ denote the flow generated by $C_H$ on $\cc$, where $s$ is the flow parameter as before. Gauge transforming a function $F$ corresponds to transporting its argument along the flow $\alpha^s_{C_H}\cdot F(x)=F(\alpha_{C_H}^s(x))$, $x\in\cc$, and reads
\begin{equation}\label{}
\alpha^s_{C_H}\cdot F(x)= \sum_{k=0}^\infty \frac{s^k}{k!} \{F,C_H\}_k(x)\,,
\end{equation}
where  $\{F,P\}_k = \{ \ldots \{\{F,P \}, P \}, \ldots, P \}$ is the $k$-nested Poisson bracket of $F$ with $C_H$. 
Clearly,
\ba
\alpha^s_{C_H}\cdot q (x)&=& q(x) + 2ps\,,\q\q\q\q\q
\alpha^s_{C_H}\cdot p (x) = p (x)\,,\nn\\
\alpha^s_{C_H}\cdot t (x)&=& t(x) + s\,,\q\q\q\q\q\,\,\,\,\,
\alpha^s_{C_H}\cdot p_t (x) = p_t (x)\,
\ea
Hence, the gauge transformation from $\cc\cap\cg_{t=0}$ to $\cc\cap\cg_{q=0}$ is
\begin{equation}\label{}
\alpha_{t \rightarrow q} := \alpha^{-q(x)/2p(x)}_{C_H}\,,
\end{equation}
i.e.\ flowing with `parameter distance' $s=-q(x)/2p(x)$, where $q(x),p(x)$ are the values of $q,p$ on the orbit prior to the transformation (of course, $p$ takes the same value on the entire orbit; it is a gauge transformation that depends on the relation between the clock and evolving degrees of freedom before the switch. This transformation is only defined for $p\neq0$. Similarly, the inverse transformation from $\cc\cap\cg_{q=0}$ to $\cc\cap\cg_{t=0}$ reads
\begin{equation}\label{}
\alpha_{q \rightarrow t} := \alpha^{-t(x)}_{C_H}\,.
\end{equation}

This finally permits us to write down the maps between the various reduced phase spaces. Firstly,
\ba
\cs_{t\to q\pm}:=\pi_{\pm}\circ\alpha_{t\to q}\circ\iota_{q|t}:\cp_{q|t}\rightarrow\cp_{\pm}\,,\nn\\
(q,p)\mapsto (t=-\f{q}{2p},\,p_t=-p^2)\,.
\ea
Notice that $q=0$ intersects of course both $\cc_+\subset\cc$ and $\cc_-\subset\cc$ (modulo the issues for $p=0$) and so $\cs_{t\to q\pm}$ indeed maps the $p<0$ part (the left moving sector) of the phase space $\cp_{q|t}$ to $\cp_+$ and the $p>0$ part (the right moving sector) of the phase space $\cp_{q|t}$ to $\cp_-$.

Conversely, we have
\ba
\cs_{q\pm\to t}:=\pi_{t=0}\circ \alpha_{q\to t}\circ\iota_\pm:\cp_\pm\rightarrow\cp_{q|t}\,,\nn\\
(t,p_t)\mapsto(q=\pm2\,t\,\sqrt{-p_t},\,p=\mp\sqrt{-p_t})\,.
\ea


\section{Consistency of the Heisenberg equations in $q$ time}\label{app_commut}

We would like to show that, as claimed in (\ref{HeisenX}),
\ba
-i\,[\hat{T}_\pm,\hat{H}_\pm]=\mp\f{1}{2}\,\widehat{(\sqrt{-p_t})_\delta^{-1}}\,.\label{C1}
\ea
To this end, notice that $[\hat{T}_\pm,\hat{H}_\pm]=\pm[\,\hat{t}_{\delta\pm},\widehat{\sqrt{-p_t}}\,]$ and
\ba
\widehat{\sqrt{-p_t}}\,[\,\hat{t}_{\delta\pm},\widehat{\sqrt{-p_t}}\,]+[\,\hat{t}_{\delta\pm},\widehat{\sqrt{-p_t}}\,]\,\widehat{\sqrt{-p_t}}=[\,\hat{p}_t,\hat{t}_{\delta\pm}\,]\underset{{\tiny(\ref{almostcan})}}{=}-i\, \widehat{(p_t)_\delta^{-1}}\,\hat{p}_t\,.\label{C2}
\ea
Next, we resort to the {\it non-regularized} $\widehat{(\sqrt{-p_t})^{-1}}$, defined through spectral decomposition. This is not a self-adjoint operator as it becomes unbounded for $p_t\rightarrow0$. However, by spectral decomposition and l'Hospital's rule it still satisfies $\widehat{(\sqrt{-p_t})^{-1}}\,\cdot\,\widehat{\sqrt{-p_t}}=\mathds{1}$ and this is all we need here. Indeed, by acting with it on (\ref{C2}) from both sides we find
\ba
[\,\hat{t}_{\delta\pm},\widehat{\sqrt{-p_t}}\,]+\widehat{(\sqrt{-p_t})^{-1}}\,[\,\hat{t}_{\delta\pm},\widehat{\sqrt{-p_t}}\,]\,\widehat{\sqrt{-p_t}}&=&i\, \widehat{(p_t)_\delta^{-1}}\,\widehat{\sqrt{-p_t}}\,\nn\\
\widehat{\sqrt{-p_t}}\,[\,\hat{t}_{\delta\pm},\widehat{\sqrt{-p_t}}\,]\,\widehat{(\sqrt{-p_t})^{-1}}+[\,\hat{t}_{\delta\pm},\widehat{\sqrt{-p_t}}\,]&=&i\, \widehat{(p_t)_\delta^{-1}}\,\widehat{\sqrt{-p_t}}\,,\label{C3}
\ea
so that
\ba
\widehat{(\sqrt{-p_t})^{-1}}\,[\,\hat{t}_{\delta\pm},\widehat{\sqrt{-p_t}}\,]\,\widehat{\sqrt{-p_t}}= \widehat{\sqrt{-p_t}}\,[\,\hat{t}_{\delta\pm},\widehat{\sqrt{-p_t}}\,]\,\widehat{(\sqrt{-p_t})^{-1}}\,.
\ea
This is only possible if $[\,\hat{t}_{\delta\pm},\widehat{\sqrt{-p_t}}\,]=f(\hat{p}_t)$. But then (\ref{C3}) implies
\ba
[\,\hat{t}_{\delta\pm},\widehat{\sqrt{-p_t}}\,]=\f{i}{2}\, \widehat{(p_t)_\delta^{-1}}\,\widehat{\sqrt{-p_t}}\,,
\ea
where $\widehat{(p_t)_\delta^{-1}}$ is defined in (\ref{invpt}). Notice that $\widehat{(p_t)_\delta^{-1}}\,\widehat{\sqrt{-p_t}}=-\widehat{(\sqrt{-p_t})_\delta^{-1}}$, where $\widehat{(\sqrt{-p_t})_\delta^{-1}}$ is defined in (\ref{invroot}). Hence, in conjunction we recover (\ref{C1}).

\section{Inverting the constraint trivialization in $q$ time}\label{app_Dirac2q}

We begin by showing that, as claimed in sec.\ \ref{sec_Dirac2q},
\ba
\ct_q^{-1}\,\ct_q=\mathds{1}\,,\q\text{on } \ch_{\rm phys}\,.\label{D1}
\ea
Indeed, using (\ref{trivialq}, \ref{trivialqinv}) we find
\ba
\ct_q^{-1}\,\ct_q&=&\exp\left(- i\,\hat{q}\,(\widehat{\sqrt{-p_t}}-\epsilon)\right)\,\theta(-\hat{p})\,\exp\left( i\,\hat{q}\,(\widehat{\sqrt{-p_t}}-\epsilon)\right)\,\theta(-\hat{p})\nn\\
&&+\exp\left(- i\,\hat{q}\,(\widehat{\sqrt{-p_t}}-\epsilon)\right)\,\theta(-\hat{p})\,\exp\left( -i\,\hat{q}\,(\widehat{\sqrt{-p_t}}-\epsilon)\right)\,\theta(\hat{p})\nn\\
&&+\exp\left( i\,\hat{q}\,(\widehat{\sqrt{-p_t}}-\epsilon)\right)\,\theta(\hat{p})\,\exp\left( i\,\hat{q}\,(\widehat{\sqrt{-p_t}}-\epsilon)\right)\,\theta(-\hat{p})\nn\\
&&+\exp\left( i\,\hat{q}\,(\widehat{\sqrt{-p_t}}-\epsilon)\right)\,\theta(\hat{p})\,\exp\left(- i\,\hat{q}\,(\widehat{\sqrt{-p_t}}-\epsilon)\right)\,\theta(\hat{p})\,.
\ea
We can now use that
\ba
\exp\left(\mp i\,\hat{q}\,(\widehat{\sqrt{-p_t}}-\epsilon)\right)\,\theta(\mp\hat{p})\,\exp\left(\pm i\,\hat{q}\,(\widehat{\sqrt{-p_t}}-\epsilon)\right)&=&\theta(\mp\hat{p}\mp\widehat{\sqrt{-p_t}}\pm\epsilon)\,,\label{D3}\\
\exp\left(\pm i\,\hat{q}\,(\widehat{\sqrt{-p_t}}-\epsilon)\right)\,\theta(\pm\hat{p})\,\exp\left(\pm i\,\hat{q}\,(\widehat{\sqrt{-p_t}}-\epsilon)\right)&=&\exp\left(\pm 2i\,\hat{q}\,(\widehat{\sqrt{-p_t}}-\epsilon)\right)\,\theta(\pm\hat{p}\mp\widehat{\sqrt{-p_t}}\pm\epsilon)\,,\nn
\ea
which can be checked by employing, e.g., the representation
\ba
\theta(\hat{p})=\int_0^{\infty}\,\int_{-\infty}^{\infty}\,\mathrm{d}p\,\mathrm{d}p_t\,\ket{p}_q\ket{p_t}_t{}_q\bra{p}{}_t\bra{p_t}\,,\q\q\q\q \theta(-\hat{p})=\int_{-\infty}^{0}\,\int_{-\infty}^{\infty}\,\mathrm{d}p\,\mathrm{d}p_t\,\ket{p}_q\ket{p_t}_t{}_q\bra{p}{}_t\bra{p_t}\,.
\ea
We thus have
\ba
\ct_q^{-1}\,\ct_q&=&\theta(-\hat{p}-\widehat{\sqrt{-p_t}}+\epsilon)\,\theta(-\hat{p})+\exp\left(- 2i\,\hat{q}\,(\widehat{\sqrt{-p_t}}-\epsilon)\right)\,\theta(-\hat{p}+\widehat{\sqrt{-p_t}}-\epsilon)\,\theta(\hat{p})\nn\\
&&+\theta(+\hat{p}+\widehat{\sqrt{-p_t}}-\epsilon)\,\theta(\hat{p})+\exp\left(+ 2i\,\hat{q}\,(\widehat{\sqrt{-p_t}}-\epsilon)\right)\,\theta(+\hat{p}-\widehat{\sqrt{-p_t}}+\epsilon)\,\theta(-\hat{p})\,.
\ea
Finally, by evaluating this last expression in the explicit representation of physical states in the form (\ref{phys2}), it is easy to convince oneself that
\ba
\ct_q^{-1}\,\ct_q\,\ket{\psi}_{\rm phys}=\Big(\theta(-\hat{p})+0+\theta(\hat{p})+0\Big)\,\ket{\psi}_{\rm phys}=\ket{\psi}_{\rm phys}\,.
\ea
Note that this last step is {\it only} possible for $\epsilon>0$ and only on $\ch_{\rm phys}$. This step also requires carefully using that $\theta(0)=\f{1}{2}$ in various instances. We have thus proven (\ref{D1}).

\section{Time-of-arrival observables under constraint trivialization}\label{app_toatrans}

We will show that the time-of-arrival quantum Dirac observables (\ref{qTX}) transform under the constraint trivialization map $\ct_q$ to the form given in (\ref{toatrans1}). For convenience, we rephrase  (\ref{qTX}) here:
\ba
\hat{T}_\delta(X):=\hat{t}_\delta+\f{1}{4}\left(\widehat{(p)_\delta^{-1}}\,(X-\hat{q})+(X-\hat{q})\,  \widehat{(p)_\delta^{-1}}\right)\,,\q\q\q\q\hat{P}_T(X):=\hat{p}_t\,.\label{E1}
\ea

Calculating the transformations of these observables under $\ct_q$ purely algebraically, i.e.\ without worrying about the states that these operators act on is rather cumbersome. We will therefore make use of a trick: given any Dirac observable $\hat{O}$ on $\ch_{\rm phys}$, $\ct_q\,\hat{O}\,\ct_q^{-1}$, is the corresponding observable on $\ch_{\rm phys}^{t|q}$. We will thus {\it only} evaluate $\ct_q\,\hat{O}\,\ct_q^{-1}$ on $\ch_{\rm phys}^{t|q}$ and this is all we care about. Thanks to (\ref{D1}), this can be done simply by first evaluating the left hand side of
\ba
\ct_q\,\hat{O}\,\ket{\psi}_{\rm phys}=\ct_q\,\hat{O}\,\ct_q^{-1}\,\ct_q\,\ket{\psi}_{\rm phys}\,,
\ea
and subsequently extracting the right hand side from it. This turns out to be much simpler.

We begin with $\hat{P}_T(X)$. Given that $\hat{p}_t$ commutes with $\ct_q$ it is obvious that
\ba
\ct_q\,\hat{P}_T(X)\,\ct_q^{-1}=\hat{p}_t\,\theta(-\hat{p})+\hat{p}_t\,\theta(\hat{p})=\hat{p}_t\,,\q\q\text{on } \ch_{\rm phys}^{t|q}\,.
\ea

We continue with $\hat{T}(X)$, which requires substantially more work.  We will divide the task into three parts (A)--(C), corresponding to the three summands in $\hat{T}(X)$ in (\ref{E1}). 

\textbf{(A)} We first compute $\ct_q\,\hat{t}_\delta\,\ct_q^{-1}$, which, in fact, can be done without the trick. We wish to compute $[\ct_q,\hat{t}_\delta]$. Since $\hat{t}_\delta$, defined in (\ref{regtime}), commutes with $\theta(\pm\hat{p})$, this step only requires 
\ba
\Big[\,\exp\left(\pm i\,\hat{q}\,(\widehat{\sqrt{-p_t}}-\epsilon)\right),\hat{t}_\delta\,\Big]=\sum_{n=0}^\infty\f{(\pm i\,\hat{q})^n}{n!}\,\Big[\,(\widehat{\sqrt{-p_t}}-\epsilon)^n,\hat{t}_\delta\Big]=\sum_{n=0}^\infty\f{(\pm i\,\hat{q})^n}{(n-1)!}\,(\widehat{\sqrt{-p_t}}-\epsilon)^{n-1}\Big[\,\widehat{\sqrt{-p_t}},\hat{t}_\delta\Big]\,.\nn
\ea
Since by (\ref{almostcan2}) $[\hat{t}_\delta,\hat{p}_t]=i\hat{p}_t \,\widehat{(p_t)^{-1}_\delta}$ we can simply repeat the algebraic steps of appendix \ref{app_commut}, which were carried out for $\hat{t}_{\delta\pm}$ of the reduced theories on $\ch_{\pm}$, but now for $\hat{t}_\delta$ defined on $\ch_{\rm kin}$, finding in complete analogy
\ba
\Big[\,\widehat{\sqrt{-p_t}},\hat{t}_\delta\,\Big]=\f{i}{2}\,\widehat{(\sqrt{-p_t})_\delta^{-1}}\,,
\ea
where $\widehat{(\sqrt{-p_t})_\delta^{-1}}$ is here the regularized Dirac observable on $\ch_{\rm phys}$ defined in (\ref{invroot2}). Hence, 
\ba
\Big[\,\exp\left(\pm i\,\hat{q}\,(\widehat{\sqrt{-p_t}}-\epsilon)\right),\hat{t}_\delta\,\Big]=\left(\mp\,\f{\hat{q}}{2}\,\widehat{(\sqrt{-p_t})_\delta^{-1}}\right)\,\exp\left(\pm i\,\hat{q}\,(\widehat{\sqrt{-p_t}}-\epsilon)\right)\,,
\ea
such that, invoking the arguments of appendix \ref{app_Dirac2q}, one finds in conjunction
\ba
\ct_q\,\hat{t}_\delta\,\ct_q^{-1}=\left(\hat{t}_\delta-\,\f{\hat{q}}{2}\,\widehat{(\sqrt{-p_t})_\delta^{-1}}\right)\,\theta(-\hat{p})+\left(\hat{t}_\delta+\,\f{\hat{q}}{2}\,\widehat{(\sqrt{-p_t})_\delta^{-1}}\right)\,\theta(\hat{p})\,,\q\q\text{on } \ch_{\rm phys}^{t|q}\,.
\ea

\textbf{(B)} Next, we compute $\ct_q\,(X-\hat{q})\,\widehat{(p)_\delta^{-1}}\,\ct_q^{-1}$ by using the above mentioned trick and evaluating $\ct_q\,(X-\hat{q})\,\widehat{(p)_\delta^{-1}}\,\ket{\psi}_{\rm phys}$ in the representation (\ref{phys2}) of physical states. Recalling the definition (\ref{invp}) of the regularized inverse $p$ operator, we firstly have
\ba
\widehat{(p)_\delta^{-1}}\,\ket{\psi}_{\rm phys}&=& \int_{-\infty}^{-\delta^2}\,\f{\mathrm{d}p_t}{2\sqrt{-p_t}}\,\Big(-\f{\psi_{\rm kin}(-\sqrt{-p_t},p_t)}{\sqrt{-p_t}}\,\ket{-\sqrt{-p_t}\,}_q\ket{p_t}_t+\f{\psi_{\rm kin}(\sqrt{-p_t},p_t)}{\sqrt{-p_t}}\,\ket{\sqrt{-p_t}\,}_q\ket{p_t}_t\Big)\nn\\
&&\!\!\!\!\!\!\!\!\!\!+\int_{-\delta^2}^0\,\f{\mathrm{d}p_t}{2\delta^2}\,\Big(-\psi_{\rm kin}(-\sqrt{-p_t},p_t)\,\ket{-\sqrt{-p_t}\,}_q\ket{p_t}_t+\psi_{\rm kin}(\sqrt{-p_t},p_t)\,\ket{\sqrt{-p_t}\,}_q\ket{p_t}_t\Big)\,.\label{invpact}
\ea
To compute $\ct_q\,\hat{q}\,\widehat{(p)_\delta^{-1}}\,\ket{\psi}_{\rm phys}$, let us check
\ba
e^{i\,\hat{q}\,(\widehat{\sqrt{-p_t}}-\epsilon)}\,\theta(-\hat{p})\,\hat{q}\,\ket{-\sqrt{-p_t}\,}_q\ket{p_t}_t&\underset{{\tiny (\ref{D3})}}{=}&\theta(-\hat{p}+\widehat{\sqrt{-p_t}}-\epsilon)\,\hat{q}\,e^{i\,\hat{q}\,(\widehat{\sqrt{-p_t}}-\epsilon)}\,\ket{-\sqrt{-p_t}\,}_q\ket{p_t}_t\nn\\
&=&\int_{-\infty}^{\sqrt{-p_t}-\epsilon}\,\mathrm{d}p\,\ket{p}_q{}_q\bra{p}\,\hat{q}\,\ket{-\epsilon}_q\ket{p_t}_t\nn\\
&=&\int_{-\infty}^{\sqrt{-p_t}-\epsilon}\,\mathrm{d}p\,\int\,\mathrm{d}q\,q\,{}_q\bra{p}q\ra\bra{q}-\epsilon\ra_q\,\ket{p}_q\ket{p_t}_t\nn\\
&=&\int_{-\infty}^{\sqrt{-p_t}-\epsilon}\,\mathrm{d}p\,\hat{q}\,\delta(p+\epsilon)\,\ket{p}_q\ket{p_t}_t\nn\\
&=&\hat{q}\,\ket{-\epsilon}_q\ket{p_t}_t\,.\label{bla1}
\ea
Similarly, one finds
\ba
e^{i\,\hat{q}\,(\widehat{\sqrt{-p_t}}-\epsilon)}\,\theta(-\hat{p})\,\hat{q}\,\ket{\sqrt{-p_t}\,}_q\ket{p_t}_t=0\,.
\ea
In fact, this last equation only holds for $p_t<0$. For $p_t=0$ it yields $\f{1}{2}\hat{q}\,\ket{-\epsilon}_q\ket{p_t}_t$, due to $\theta(0)=\f{1}{2}$, just like actually (\ref{bla1}) does for $p_t=0$.

In complete analogy, one can check that 
\ba
e^{-i\,\hat{q}\,(\widehat{\sqrt{-p_t}}-\epsilon)}\,\theta(-\hat{p})\,\hat{q}\,\ket{\pm\sqrt{-p_t}\,}_q\ket{p_t}_t=\f{1\pm1}{2}\hat{q} \, \ket{\epsilon\,}_q\ket{p_t}_t\,
\ea
(except that at $p_t=0$ one again actually has $\f{1}{2}\hat{q}\,\ket{\epsilon}_q\ket{p_t}_t$ for both cases).

Combining all these results, it is straightforward to convince oneself that
\ba
\ct_q\,(X-\hat{q})\,\widehat{(p)_\delta^{-1}}\,\ket{\psi}_{\rm phys}=\Big[(\hat{q}-X)\,\widehat{(\sqrt{-p_t})_\delta^{-1}}\,\theta(-\hat{p})-(\hat{q}-X)\,\widehat{(\sqrt{-p_t})_\delta^{-1}}\,\theta(\hat{p})\Big]\,\ct_q\,\ket{\psi}_{\rm phys}\,,
\ea
where $\widehat{(\sqrt{-p_t})_\delta^{-1}}$ is again given by (\ref{invroot2}). It appears here because, up to sign in the left moving sector, it has the same action on $\ket{\psi}_{\rm phys}$, as $\widehat{(p)^{-1}_\delta}$ does in (\ref{invpact}). This is also the reason for the crucial sign difference between the left and right moving sector in the last expression. Altogether, this yields
\ba
\ct_q\,(X-\hat{q})\,\widehat{(p)_\delta^{-1}}\,\ct_q^{-1}=(\hat{q}-X)\,\widehat{(\sqrt{-p_t})_\delta^{-1}}\,\theta(-\hat{p})-(\hat{q}-X)\,\widehat{(\sqrt{-p_t})_\delta^{-1}}\,\theta(\hat{p})\,,\q\q\text{on } \ch_{\rm phys}^{t|q}\,.
\ea

\textbf{(C)} We proceed with computing $\ct_q\,\widehat{(p)_\delta^{-1}}\,(X-\hat{q})\,\ct_q^{-1}$, again, by evaluating $\ct_q\,\widehat{(p)_\delta^{-1}}\,(X-\hat{q})\,\ket{\psi}_{\rm phys}$. To this end, recalling the definition (\ref{invp}) of the inverse $p$ operator, it is useful to single out
\ba
\widehat{(p)_\delta^{-1}}\,\hat{q}\,\ket{\mp\sqrt{-p_t}\,}_q\ket{p_t}_t&=&\f{1}{\sqrt{2\pi}}\int_{-\infty}^\infty\,\mathrm{d}q\,q\,e^{\mp i\,q\,\sqrt{-p_t}}\,\widehat{(p)_\delta^{-1}}\,\ket{q}_q\ket{p_t}_t\nn\\
&=&\left[\int_{-\infty}^{-\delta}\,\f{\mathrm{d}p}{p}\,\hat{q}\,\delta(p\pm\sqrt{-p_t})+\int_{-\delta}^{+\delta}\,\f{\mathrm{d}p\,p}{\delta^2}\,\hat{q}\,\delta(p\pm\sqrt{-p_t})\right.\nn\\
&&\left.\q\q\q\q\q\q\q\q\q+\int_{+\delta}^{+\infty}\,\f{\mathrm{d}p}{p}\,\hat{q}\,\delta(p\pm\sqrt{-p_t})\right]\,\ket{p}_q\ket{p_t}_t\nn\\
&=&\left[\hat{q}\,\widehat{(p)_\delta^{-1}}\,+i\,\widehat{(p)_\delta^{-2}}\,\right]\,\ket{\mp\sqrt{-p_t}\,}_q\ket{p_t}_t\,,\label{bla2}
\ea
where from the first to the second line we have inserted a completeness relation $\mathds{1}_q=\int\,\mathrm{d}p\,\ket{p}_q{}_q\bra{p}$. From the second to the third line we have used that $\hat{q}$ acts as a derivative operator, and in this last step we have defined the regularized inverse square power of $p$ as follows:
\ba\label{invp2}
\widehat{(p)_\delta^{-2}}\,\ket{p}_q:=\begin{cases}
  \f{1}{p^2}\,\ket{p}_q    & |p|\geq\delta, \\
  \f{1}{\delta^2}\,\ket{p}_q    & |p|<\delta\,.
\end{cases}
\ea

Notice that the $\hat{q}\,\widehat{(p)_\delta^{-1}}$ term in (\ref{bla2}) will lead to exactly the same result as (B). On the other hand, comparing (\ref{invp2}) with (\ref{invpt2}) and using the representation of physical states (\ref{phys2}), it is easy to see that
\ba
\ct_q\,i\,\widehat{(p)_\delta^{-2}}\,\ket{\psi}_{\rm phys}=-\ct_q\,i\,\widehat{(p_t)_\delta^{-1}}\,\ket{\psi}_{\rm phys}=-i\,\widehat{(p_t)_\delta^{-1}}\,\ct_q\,\ket{\psi}_{\rm phys}\,.
\ea

Hence, together we have 
\ba
\ct_q\,\widehat{(p)_\delta^{-1}}\,(X-\hat{q})\,\ct_q^{-1}=(\hat{q}-X)\,\widehat{(\sqrt{-p_t})_\delta^{-1}}\,\theta(-\hat{p})-(\hat{q}-X)\,\widehat{(\sqrt{-p_t})_\delta^{-1}}\,\theta(\hat{p})+i\,\widehat{(p_t)_\delta^{-1}}\,,\,\,\,\text{on } \ch_{\rm phys}^{t|q}\,.
\ea

Combining now (A)--(C), we finally find on $\ch_{\rm phys}^{t|q}$
\ba
\ct_q\,\hat{T}_\delta(X)\,\ct_q^{-1}&=&\left(\hat{t}_\delta-\f{X}{2}\,\widehat{(\sqrt{-p_t})_\delta^{-1}}\right)\,\theta(-\hat{p})+\left(\hat{t}_\delta+\f{X}{2}\,\widehat{(\sqrt{-p_t})_\delta^{-1}}\right)\,\theta(\hat{p})+\f{i}{4}\widehat{(p_t)_\delta^{-1}}\,,\label{E17}
\ea
as claimed in (\ref{toatrans1}) in sec.\ \ref{sec_Dirac2q}.

\section{Recovering the reduced evolving observables in $q$ time}\label{app_recover}

In this appendix, we prove that, as claimed in (\ref{toatrans2}) (the second equation in (\ref{toatrans2}) is evident):
\ba
\widehat{(-p_t)^{1/4}}\,\ct_q\,\hat{T}_\delta(X)\,\ct_q^{-1}\,\widehat{(-p_t)^{-1/4}}&=&\left(\hat{t}_{\delta+}-\f{X}{2}\,\widehat{(\sqrt{-p_t})_\delta^{-1}}\right)\,\theta(-\hat{p})+\left(\hat{t}_{\delta-}+\f{X}{2}\,\widehat{(\sqrt{-p_t})_\delta^{-1}}\right)\,\theta(\hat{p})\,.\nn
\ea

Given that the two definitions of $\widehat{(\sqrt{-p_t})_\delta^{-1}}$ on $\ch_\pm$ in (\ref{invpt}) and on $\ch_{\rm kin}$ in (\ref{invpt2}) coincide upon identifying $\ket{p_t}_\pm=\sqrt{-p_t}\,\ket{p_t}_t$ (as done in (\ref{lasttrans})) and commute with $\widehat{(-p_t)^{1/4}}$, we only have to check the transformation of $\hat{t}_\delta$ in (\ref{E17}). This is an operator (\ref{regtime}) defined on $\ch_{\rm kin}$, while $\hat{t}_{\delta\pm}$ above is defined on $\ch_\pm$ in (\ref{regtime0}). Working directly in momentum representation and recalling the definition (\ref{regtime}), we find
\ba
\widehat{(-p_t)^{1/4}}\,\hat{t}_\delta\,\widehat{(-p_t)^{-1/4}}\!\!\!\!\!&=&\!\!\!\!\begin{cases}
 (-p_t)^{1/4}\,i\,\p_{p_t}\, (-p_t)^{-1/4} = i\,\p_{p_t}-\f{i}{4p_t} \,,   & \q\,\,\, p_t\leq-\delta^2, \\
  (-p_t)^{1/4}\,\left(-\f{i\,p_t}{\delta^2}\,\p_{p_t}-\f{i}{2\delta^2}\right)\, (-p_t)^{-1/4} = -\f{i\,p_t}{\delta^2}\,\p_{p_t}-\f{i}{2\delta^2}+\f{i}{4\delta^2} \, ,  & -\delta^2<p_t\leq0\,.
\end{cases}\nn\\
&=&\hat{t}_\delta-\f{i}{4}\,\widehat{(p_t)^{-1}_\delta}\,.
\ea
Upon this transformation, we thus see that the last term in (\ref{E17}) gets cancelled and we have 
\ba
\widehat{(-p_t)^{1/4}}\,\ct_q\,\hat{T}_\delta(X)\,\ct_q^{-1}\,\widehat{(-p_t)^{-1/4}}&=&\left(\hat{t}_{\delta}-\f{X}{2}\,\widehat{(\sqrt{-p_t})_\delta^{-1}}\right)\,\theta(-\hat{p})+\left(\hat{t}_{\delta}+\f{X}{2}\,\widehat{(\sqrt{-p_t})_\delta^{-1}}\right)\,\theta(\hat{p})\,.\nn
\ea

Notice that $\hat{t}_\delta$ still appears in this expression, rather than $\hat{t}_{\delta\pm}$. Since this observable now acts on states of the form (\ref{lasttrans}), it thus remains to be shown that $\hat{t}_\delta\,\ket{\psi}_\pm=\hat{t}_{\delta\pm}\,\ket{\psi}_\pm$. We continue to work in momentum representation from above
\ba
\hat{t}_\delta\,\ket{\psi}_\pm&=&\hat{t}_\delta\int_{-\infty}^0\,\f{\mathrm{d}p_t}{-p_t}\,\psi_\pm(p_t)\,\ket{\mp\epsilon}_q\ket{p_t}_\pm=\hat{t}_\delta\int_{-\infty}^0\,\f{\mathrm{d}p_t}{\sqrt{-p_t}}\,\psi_\pm(p_t)\,\ket{\mp\epsilon}_q\ket{p_t}_t \nn\\
&=& \int_{-\infty}^{-\delta^2}\,{\mathrm{d}p_t}\,i\,\p_{p_t}\,\f{\psi_\pm(p_t)}{\sqrt{-p_t}}\,\ket{\mp\epsilon}_q\ket{p_t}_t +\int_{-\delta^2}^{\infty}\,{\mathrm{d}p_t}\,\left(-\f{i\,p_t}{\delta^2}\,\p_{p_t}-\f{i}{2\delta^2}\right)\,\theta(-p_t)\,\f{\psi_\pm(p_t)}{\sqrt{-p_t}}\,\ket{\mp\epsilon}_q\ket{p_t}_t \nn\\
&=&\int_{-\infty}^{-\delta^2}\f{\mathrm{d}p_t}{-p_t}\left(i\,\p_{p_t}-\f{i}{2p_t}\right){\psi_\pm(p_t)}\ket{\mp\epsilon}_q\ket{p_t}_\pm +\int_{-\delta^2}^{\infty}\f{\mathrm{d}p_t}{-p_t}\,\left(-\f{i\,p_t}{\delta^2}\p_{p_t}\right)\theta(-p_t){\psi_\pm(p_t)}\ket{\mp\epsilon}_q\ket{p_t}_\pm.\nn
\ea
From the first to the second line we have used that $\hat{t}_\delta$ is defined on $\ch_{\rm kin}$ and thus a derivative operator with respect to the standard Lebesgue measure. Since this requires letting the total integral run over all of $\mathbb{R}$, we have introduced the theta function to cut off the integral. Its derivative in the last term will yield a delta function, which, however, can be dropped as it only `clicks' at $p_t=0$ and we have $\psi_\pm(0)=0$ assuming $\ket{\psi}_\pm$ lies in the domain of $\hat t_{\delta\pm}$. We may thus henceforth drop the theta function in the last term if we simultaneously restrict the integral to run only from $-\delta^2$ to $0$.

Recall now from (\ref{regtime0}) that
\ba
\hat{t}_{\delta\pm}=\f{1}{2}\left(\widehat{(p_t)_\delta^{-1}}\,\hat{\ft}+\hat{\ft}\,\widehat{(p_t)_\delta^{-1}}\right)\,,
\ea
which, using the affine momentum representation (\ref{affrep}), becomes
\ba
\hat{t}_{\delta\pm}=\begin{cases}
 \f{i}{2}\left(\p_{p_t}+p_t\,\p_{p_t}\f{1}{p_t}\right) = i\,\p_{p_t} - \f{i}{2p_t}\,,     & \q\q p_t\leq-\delta^2, \\
-\f{i\,p_t}{\delta^2}\,\p_{p_t}\,,& -\delta^2<p_t\leq0\,.
\end{cases}
\ea
Hence, we indeed find
\ba
\hat{t}_\delta\,\ket{\psi}_\pm=\hat{t}_{\delta\pm}\,\ket{\psi}_\pm\,,
\ea
thus ultimately proving the claim (\ref{toatrans2}).

\section{Switching between $t$ and $q$ time in the quantum theory}\label{app_t2q}

It is straightforward to verify the claim of sec.\ \ref{sec_qclockswitch}. We had
\ba
\hat{\cs}_{t\to q\pm}\,\ket{\psi}_{q|t}=2\sqrt{\pi}\,{}_q\bra{q=0}\,\theta(\mp\hat{p})\,(-\hat{p}_t)^{1/4}\,\ct_q\,\ct_t^\dag\,\ket{p_t=0}_t\otimes\,\,\ket{\psi}_{q|t}=\f{1}{\sqrt{2}}\,\ket{\psi}_\pm\,. \label{G1}
\ea
We can now simplify this map  considerably, using only properties of the reduced Hilbert spaces. To this end, we perform the variable transformations $p=-\sqrt{-p_t}$ for $p<0$ and $p=\sqrt{-p_t}$ for $p>0$ and write
\ba
\ket{\psi}_{q|t}=\int_{-\infty}^\infty\,\mathrm{d}p\,\psi_{q|t}(p)\,\ket{p}_q=\int_{-\infty}^0\,\f{\mathrm{d}p_t}{2\sqrt{-p_t}}\,\Big(\psi_{\rm kin}(-\sqrt{-p_t},p_t)\,\ket{-\sqrt{-p_t}\,}_q+\psi_{\rm kin}(\sqrt{-p_t},p_t)\,\ket{\sqrt{-p_t}\,}_q\Big)\,.\nn
\ea
Recalling ${}_q\bra{p}p'\ra_q=\delta(p-p')$, while ${}_\pm\bra{p_t}p'_t\ra_\pm=-p_t\,\delta(p_t-p_t')$, it is now straightforward to check that
\ba
\sqrt{2}\,\hat{\cp}_{q\to t}\,\theta(\mp\hat{p})\,\widehat{\sqrt{|p|}}\,\ket{\psi}_{q|t}=\f{1}{\sqrt{2}}\,\ket{\psi}_\pm\,,
\ea
where, in some analogy to the parity-swap operator of \cite{Giacomini:2017zju} (see also \cite{Vanrietvelde:2018pgb}), we have defined
\ba
\hat{\cp}_{q\to t}\,\ket{p}_q:=\f{1}{|p|}\,\ket{-p^2}_\pm\,.\label{swap}
\ea
We thus have the equivalence
$
\hat{\cs}_{t\to q\pm} \equiv \sqrt{2}\,\hat{\cp}_{q\to t}\,\theta(\mp\hat{p})\,\widehat{\sqrt{|p|}}\,.
$
The right hand side is now only expressed in terms of structures from the reduced Hilbert spaces.

Conversely, we had
\ba
\hat{\cs}_{q\pm\to t}\,\ket{\psi}_\pm&=&\theta(\mp\hat{p})\,\ket{\psi}_{q|t}\,.
\ea
This map can also be simplified and expressed entirely in terms of properties of the reduced Hilbert spaces. Indeed, using the same variable transformations as above, it is easy to check that
\ba
\hat{\cs}_{q\pm\to t}\equiv \f{1}{2}\,\hat{\cp}_{t\to q\pm}\,\widehat{(-p_t)^{-1/4}}\,,
\ea
where, in analogy to (\ref{swap}), we have defined the swap
\ba
\hat{\cp}_{t\to q\pm}\,\ket{p_t}_\pm:=\sqrt{-p_t}\,\ket{\mp\sqrt{-p_t}\,}_q\,.
\ea

\providecommand{\href}[2]{#2}\begingroup\raggedright\endgroup

\end{document}